\documentclass[aip, amsmath,amssymb,reprint]{revtex4-2}
\usepackage[utf8]{inputenc}
\usepackage[T1]{fontenc}
\usepackage{graphicx}
\usepackage{dcolumn, lipsum,graphicx,float}
\usepackage{diagbox}
\graphicspath{ {./Figures/} }
\usepackage{slashbox}
\usepackage{gensymb}
\usepackage{wrapfig}
\usepackage[dvipsnames]{xcolor}
\usepackage{hyperref}
\hypersetup{
    colorlinks=true,
    linkcolor=blue,
    citecolor = blue,
    urlcolor=blue
}

\allowdisplaybreaks[4]
\usepackage{subfigure}
\usepackage{xcolor}
\usepackage{soul}
\soulregister\cite7
\soulregister\ref7

\begin{document}

\title{Quantum heat valve and entanglement in superconducting $L C$ resonators}

\author{Yu-qiang Liu}
\affiliation{School of Physics, Dalian University of Technology, Dalian 116024, China}
\author{ Yi-jia Yang}
\affiliation{School of Physics, Dalian University of Technology, Dalian 116024, China}
\author{Ting-ting Ma}
\affiliation{School of Physics, Dalian University of Technology, Dalian 116024, China}
\author{Chang-shui Yu}
\email{ycs@dlut.edu.cn}

\affiliation{School of Physics, Dalian University of Technology, Dalian 116024, China}

\date{\today}


\begin{abstract}
Quantum superconducting circuit with flexible coupler has been a powerful platform for designing quantum thermal machines. In this letter, we employ the tunable coupling of two superconducting resonators to realize a heat valve by modulating magnetic flux using
a superconducting quantum interference device (SQUID). It is shown that a heat valve can be realized in a wide parameter range.  We find a consistent relation between the heat current and quantum entanglement, which indicates the dominant role of entanglement on the heat valve.  It provides an insightful understanding of quantum features in quantum heat machines. 
\end{abstract}

\maketitle


In recent years, the rapid development in quantum thermodynamics \cite{vinjanampathy2016quantum} and condensed matter systems \cite{RevModPhys.93.041001} has deepened our understanding of them. One of the key topics of quantum thermodynamics is the manipulation of quantum heat transport. So far, quantum heat devices have been well developed based on different platforms, including atomic systems \cite{PhysRevLett.116.200601, PhysRevE.89.062109, PhysRevE.95.022128, PhysRevE.99.042121, PhysRevE.99.032112, PhysRevE.106.024110}, quantum dots \cite{PhysRevB.101.075417}, harmonic oscillators \cite{PhysRevE.85.061126}, and quantum superconducting circuits \cite{ PhysRevB.94.235420, tan2017quantum,  gubaydullin2022photonic, PhysRevE.102.030102, PhysRevLett.119.090603, PhysRevE.107.044121}. The superconducting circuit is undoubtedly of particular concern due to its flexible controllability and feasible experiment implementation. In particular, the internal coupling of the superconducting quantum circuit has entered the ultra-strong, even deep-strong coupling regime \cite{RevModPhys.91.025005, RevModPhys.93.025005, PhysRevA.98.053859, PhysRevA.95.053824, PhysRevA.84.043832, frisk2019ultrastrong, PhysRevApplied.16.044045} and has extensively explored and produced novel applications including optical phenomena, and quantum simulation \cite{niemczyk2010circuit, frisk2019ultrastrong}. Quantum superconducting circuit system has been used to design quantum thermodynamic devices with some functions such as refrigeration \cite{tan2017quantum, PhysRevB.96.115408, PhysRevB.94.235420},  rectification \cite{ronzani2018tunable, PhysRevApplied.15.054050, Karimi_2017}, amplification \cite{PhysRevB.101.184510}, and measurement of temperature \cite{PhysRevLett.119.090603}, etc. It is shown that quantum thermal valves,  implementing heat modulation, have been realized by a superconducting qubit system coupled to heat reservoirs \cite{PhysRevB.103.104304, ronzani2018tunable}.  The superconducting quantum circuit has become a powerful platform for various quantum thermal machines.

The usual coupling of superconducting circuit components is based on either a fixed or a tunable coupler. Tunable couplers \cite{Tian_2008, PhysRevLett.104.177004, PhysRevLett.106.060501, PhysRevApplied.15.064074, Brink_2005} have been an efficient method for controlling the coupling energy via a superconducting quantum interference device (SQUID). For example, Ref. \cite{PhysRevA.105.052605} designs a flux-tunable transverse and longitudinal coupling. Recently, the authors have implemented an inter-resonator ultrastrong coupler, and the coupling strength has wide-range tunability involving antiferromagnetic and ferromagnetics from $-1086$ to $604 \mathrm{MHz}$ \cite{PhysRevApplied.16.064041}.
Inspired by Ref. \cite{PhysRevApplied.16.064041}, we will employ the tunable coupling superconducting resonators to modulate heat transport via external magnetic flux, which extends Ref. \cite{PhysRevApplied.16.064041} to the open quantum system with two heat reservoirs at different temperatures. Moreover, what role do quantum features, such as entanglement, coherence, purity, etc., play in realizing a quantum thermodynamic function? This question has attracted increasing interest, but the answer could strongly depend on the particular system and thermodynamic function to be realized \cite{vinjanampathy2016quantum, Goold_2016, PhysRevA.105.052605, PhysRevX.5.031044, Bohr_Brask_2015, Khandelwal_2020, Hammam_2022}. In this Letter, we design a quantum heat valve to modulate the heat transport by external magnetic flux. Meanwhile, we will find out the potential relation between heat modulation and quantum entanglement.

\begin{figure}
\includegraphics[width=90 mm]{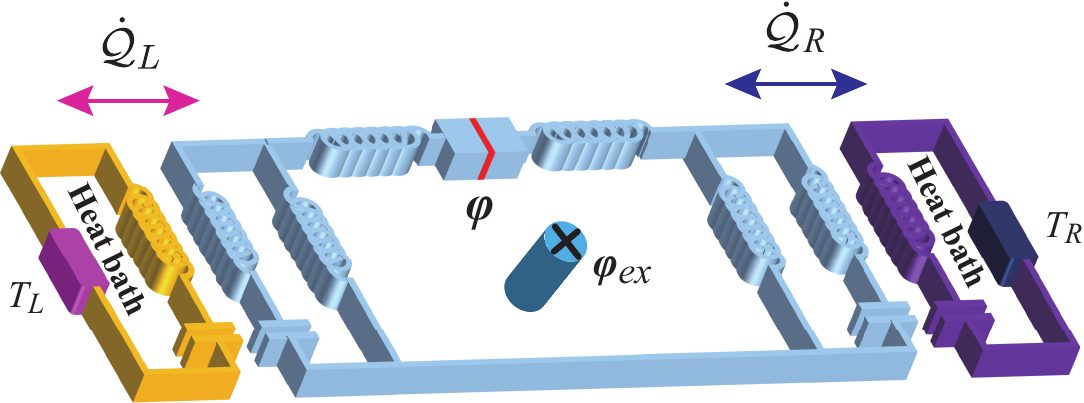}
\caption{\label{fig: model} Sketch of quantum heat valve consisting of two inductive resonators coupled to heat baths by mutual inductance. Each heat reservoir consists of a resistor $R_{\nu}$, an inductor $L_{\nu}$, and a capacitor $C_{\nu}$ (RLC) in the series circuit at different temperature $T_{\nu}$, and $\dot{\mathcal{Q}_{\nu}}$ reflect heat flows between system and heat reservoir.  
}
\end{figure}

Our system consists of ultrastrong coupling superconducting resonators, as shown in Fig. \ref{fig: model}.  The bare frequencies and coupling strength of two resonators can be flexibly modulated.  
The Hamiltonian of two coupling resonators reads ($\hbar=1$ and $k_{B}=1$)
\begin{align} \label{H_{S}}
H_{S} &=\omega_{L} a^{\dagger} a+\omega_{R} b^{\dagger} b-g (a^{\dagger}-a)(b^{\dagger}-b),
\end{align}
where $\omega_{\nu}(\varphi_{ex})=\frac{1}{\sqrt{L_{m \nu} C_{\nu}}}$ with $\nu=L, R$ and $g (\varphi_{ex})=\sqrt{Z_{L} Z_{R}}/(2 M_{m})$ denote the bare frequencies and interaction strength for resonators, respectively. To realize a flux-modulated system, we have applied a magnetic field perpendicular to the circuit controlled by a direct current coil which brings an external magnetic flux $\varphi_{ex}$ into the SQUID \cite{PhysRevB.96.174520, sandberg2008tuning, Uhl}.

Let each resonator couple with an RLC circuit  as Ref. \cite{PhysRevLett.100.155902} serving as the heat reservoir with its free Hamiltonian given as
$
H_{\nu}=\sum_{ k}\omega_{ k}{b}_{\nu k}^{\dagger}{b}_{\nu k},
$
where $b_{\nu k}$, $\omega_{\nu k}$ denote the annihilation operator and frequency of the $k$th mode of the reservoir $\nu$. The annihilation and creation operators satisfy the relation $[b_{\nu l}, b^{\dagger}_{\mu k}]=\delta_{\nu \mu, lk}$.  The interaction Hamiltonian of the resonator and the reservoir is given by \cite{PhysRevLett.100.155902,caldeira1983quantum, PhysRevB.81.144510, PhysRevA.89.023817}
\begin{equation}
H_{I} =(a+a^{\dagger}) (B^{\dagger}_{L}+B_{L})+(b+b^{\dagger}) (B^{\dagger}_{R}+B_{R}),\label{se}
\end{equation}
where we have set $B_{\nu}=\sum_{k}\kappa_{\nu k} b_{\nu k}$ and $\kappa_{\nu k}$ denote the system-reservoir coupling strength. The noise spectrum is $\Gamma_{\nu}(\omega)=2 \omega \mathrm{Re} Y_{\nu}(\omega)N^{\nu}$ with $\mathrm{Re} Y_{\nu}(\omega)=\frac{1}{R_{\nu} [1+Q_{\nu}^{2}(\frac{\omega}{\omega_{LC \nu}}-\frac{\omega_{LC \nu}}{\omega})^2]}$ \cite{PhysRevLett.100.155902, PhysRevB.76.174523} and quality factor of LC circuit $Q_{\nu}=\frac{1}{R_{\nu}}\sqrt{L_{\nu}/C_{\nu}}$.

The evolution of the coupled-resonator system is governed by the global master equation \cite{breuer2002theory, PhysRevA.84.043832} based on the Born-Markov-Secular approximation.  The explicit form of the master equation reads
\begin{equation}
\begin{split}
 \label{Eq. ms}
\frac{d \rho}{dt}&=\sum_{j=\pm}\lbrace
(\Gamma_{R} (\omega_{j}) |X_{j}|^{2} +\Gamma_{L}(\omega_{j}) |W_{j}|^{2}) \mathcal{D}[p^{\dagger}_{j}] \rho \\&+(\Gamma_{L} (-\omega_{j}) |W_{j}|^{2}+\Gamma_{R} (-\omega_{j}) |X_{j}|^{2} )\mathcal{D}[p_{j}] \rho \rbrace,
\end{split}
\end{equation}
where the dissipator $\mathcal{D}[R] \rho=R \rho R^{\dagger}-\frac{1}{2}\left\{R^{\dagger} R, \rho\right\}$,    $p_{\pm}=w_{\pm} a+x_{\pm} b+y_{\pm} a^{\dagger}+z_{\pm} b^{\dagger}$ denote the eigenmode of the Hamiltonian (\ref{H_{S}}) with $[p_{i}, p^{\dagger}_{j}]=\delta_{i, j}$ and the normalized coefficients $w_{\pm}$, $y_{\pm}$, $x_{\pm}$, and $z_{\pm}$ , and $W_{j}=(w_j-y_j)$ and $X_{j}=(x_j -z_j)$ (Supplementary Material).
The spectral densities are $\Gamma_{L} (\omega_{\pm})=\frac{1}{2} G_{L} (\omega_{\pm})=\zeta^{L}_{\pm} N^{L} (\omega_{\pm})$ and $\Gamma_{R} (\omega_{\pm})=\frac{1}{2} G_{R} (\omega_{\pm})=\zeta^{R}_{\pm} N^{R} (\omega_{\pm})$ shown in Eq. (S9) of Supplementary Material, and the mean photon number $N^{\nu} (\omega_{\pm})=\frac{1}{e^{ \omega_{\pm}/ T_{\nu}}-1}$. 
Obviously,  $\zeta^{\nu}_{\pm}=2 \omega \mathrm{Re} Y_{\nu}(\omega)$ and $N^{\nu}_{\pm}=N^{\nu} (\omega_{\pm})$. 
The detailed derivation of the master equation is given in  Supplementary Material. In addition, we would like to emphasize that many methods can be used to realize the heat reservoir, which will provide different spectral densities \cite{caldeira1983quantum, vool2017introduction, https://doi.org/10.1002/qute.202100054}.

To describe the thermal transport of the open quantum system, we will address the steady-state heat current \cite{breuer2002theory} defined by 
\begin{equation} \label{heatcurrent}
\dot{\mathcal{Q}}_{\nu}=\operatorname{Tr}\left\lbrace H_{S}\mathcal{L}_{\nu} (\rho_{ss})\right\rbrace , 
\end{equation}
where $\dot{\mathcal{Q}}_{\nu}>0$ denotes the heat transfer from the reservoir into the system and $\dot{\mathcal{Q}}_{\nu}<0$ denotes the heat from the system into the  reservoir,  and $\rho_{ss}$ 
is the steady state solving from $\frac{d \rho}{dt}=0$. 
In the current open system,  one can find the heat current $\dot{\mathcal{Q}}_{L}$ reads
\begin{align}
\nonumber \dot{\mathcal{Q}}_{L}=&|W_{+}|^{2} \zeta^{L}_{+} (N^{L}_{+}-\langle p^{\dagger}_{+} p_{+}\rangle )+|W_{-}|^{2} \zeta^{L}_{-} (N^{L}_{-}-\langle p^{\dagger}_{-} p_{-} \rangle)\\ 
=&\sum_{j=\pm}\frac{(N^{L}_{j}-N^{R}_{j}) |W_{j}|^{2} |X_{j}|^{2} \zeta^{L}_{j} \zeta^{R}_{j} \omega_{j}}{|W_{j}|^{2} \zeta^{L}_{k}+|X_{j}|^{2} \zeta^{R}_{j}}.\label{hc dd}
\end{align}
Here, we assume $T_L>T_R$, heat current flows from the left (hot) reservoir to the right (cold) reservoir shown in Fig. \ref{fig: model}.
\begin{figure}
  \centering
\includegraphics[width=85 mm]{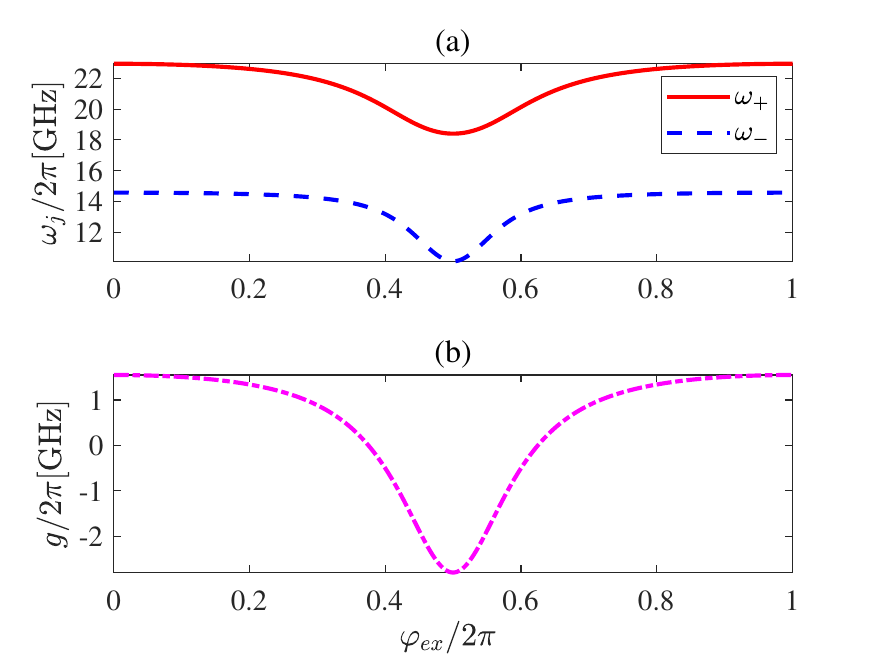}
\caption{The polaritonic frequencies $\omega_{\pm}$ (a) and coupling strength $g$ (b) of resonators versus external magnetic flux. The related parameters are considered: the self-inductance $L_a=2.023 \mathrm{nH}$, $L_b=2.023 \mathrm{nH}$, the capacitance of resonator A $C_a= 42.3 \mathrm{fF}$,
the capacitance of resonator B $C_b=18.27\mathrm{fF}$,
shared inductance $L_{sh}=0.446 \mathrm{nH}$,
junction inductance $L_{J 0}=1.210 \mathrm{nH}$,
geometric mutual inductance $M_{0}=0.381 \mathrm{nH}$,
unshared loop inductance $L_{0}= 0.177 \mathrm{nH}$,
offset of inductance modulation $\delta=0.053$.}\label{fig: mod_omega_g}
\end{figure}

To study the entanglement of the system, we will consider the quadrature field operators of the system $\lbrace d_{\pm}, f_{\pm}\rbrace$ defined by \begin{equation}
 d_{\pm}=\frac{p_{\pm}^{\dagger}+p_{\pm}}{\sqrt{2}}, f_{\pm}=i\frac{p_{\pm}^{\dagger}-p_{\pm}}{\sqrt{2}}.
\end{equation}
The quadrature operators can be formally arranged in the vector  $\mathbf{r}:=(d_{+}, f_{+}, d_{-}, f_{-} )^{T}$, which is related to the original $\mathbf{s}=(x_{L}, y_L, x_R, y_R)^{T}$ by $\mathbf{s}=S \mathbf{r}$ with the transformation $S$ given in Eq. (S15) of Supplementary Material.

Define $\Gamma_{i, j}=\langle \lbrace \Delta \xi_{j}, \Delta \xi_{i}\rbrace\rangle=\langle \frac{\Delta \xi_{i} \Delta \xi_{j}+\Delta \xi_{j} \xi_{i}}{2}\rangle$
with $\Delta \xi=\xi_{i}-\langle \xi_{j}\rangle$ for the vector $\mathbf{r}$, we have
\begin{align} \label{Cov Matrix}
\Gamma=\left(\begin{array}{cccc}
a & 0 & 0 & 0 \\
0 & a & 0 & 0 \\
0 & 0 & b & 0 \\
0 & 0 & 0 & b
\end{array}\right),
\end{align}
where $a=\left\langle d_+^2\right\rangle$, and $b=\left\langle d_-^2\right\rangle$, given in Eq. (S14) of Supplementary Material. From Eq. (12) of Supplementary Material, one can find that $a=\left\langle p^{\dagger}_{+} p_{+}\right\rangle+ \frac{1}{2}$ and $b=\left\langle p^{\dagger}_{-} p_{-}\right\rangle+ \frac{1}{2}$. Transforming it into the original picture by the transformation $S$,  $\Gamma$ can be given by 
\begin{align}
\Gamma^{\prime}=S \Gamma S^{T}=\left(\begin{array}{cccc}
M & K  \\
K^{T} & N 
\end{array}\right),
\end{align}
where $M=M^{T}$, $N=N^{T}$ and $K$ are $2\times2$ real matrices.
\begin{figure}
\includegraphics[width=85 mm]{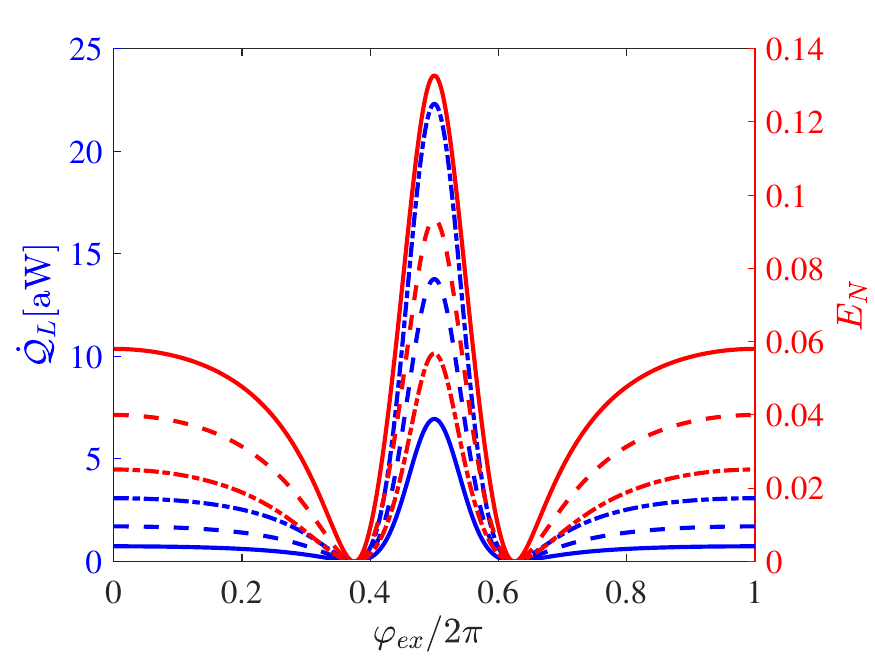}
\caption{\label{fig: Heat_EN} Heat currents and entanglement dependent on the external magnetic flux $\varphi_{ex}$   for different temperatures $T_{L}$. The solid, dashed, dot-dashed, and dotted lines denote $T_{L}=0.2 \mathrm{K}$, $0.25 \mathrm{K}$, and $0.3 \mathrm{K}$, respectively. The parameters: $R_{L}=R_{R}=1 \Omega$, $Q_{L}=Q_{R}=5$, $\omega_{LCL}=\omega_{LCR}=10 \mathrm{GHz}$, $T_{R}=0.1 \mathrm{K}$, and other parameters are the same in Fig. \ref{fig: mod_omega_g}.}
\end{figure}

Since the system is a two-mode Gaussian state, the entanglement $E_N$ \cite{PhysRevA.65.032314, PhysRevA.70.022318} can be measured by 
\begin{align}
&E_N=\mathrm{max}\left[ 0, -\mathrm{ln}(2 \nu) \right],\\
&\nu=\sqrt{\frac{\Delta-\sqrt{\Delta^2-4 \mathrm{Det} (\Gamma^{\prime})}}{2}},
\end{align}
where $\Delta=\mathrm{Det} (M)+\mathrm{Det} (N)-2\mathrm{Det} (K)$.

To numerically study the heat currents and entanglement, we take the following parameters in Ref. \cite{PhysRevApplied.16.064041}: the self-inductance $L_a=2.023 \mathrm{nH}$, $L_b=2.023 \mathrm{nH}$, the capacitance of resonator A $C_a= 42.3 \mathrm{fF}$,
the capacitance of resonator B $C_b=18.27\mathrm{fF}$,
shared inductance $L_{sh}=0.446 \mathrm{nH}$,
junction inductance $L_{J 0}=1.210 \mathrm{nH}$,
geometric mutual inductance $M_{0}=0.381 \mathrm{nH}$,
unshared loop inductance $L_{0}= 0.177 \mathrm{nH}$,
offset of inductance modulation $\delta=0.053$. The polaritonic frequencies and interaction strength of two resonators can be tunable via the external flux as the magnetic flux through the SQUID loop \cite{PhysRevApplied.16.064041}. Their dependence relations are displayed in Fig. \ref{fig: mod_omega_g}. Since $L_{J}(\varphi)=L_{J0}/(\cos(\varphi) -\delta)+L_{0}$ depends on the phase difference at the two ends of the junction and has a period of $2 \pi$, one can focus on modulation in a period. Fig. \ref{fig: mod_omega_g} indicates that the spectrum $\omega_{\pm}$ and interaction strength $g$ of the coupled resonators can vary continuously, and $g$ changes from negative to positive, going smoothly through zero at $\varphi_{ex}/2 \pi\simeq 0.374$ and $\varphi_{ex}/2 \pi \simeq 0.624$.  

In Fig. \ref{fig: Heat_EN}, we plot the heat current $\dot{\mathcal{Q}}_{L}$ versus the magnetic flux $\varphi_{ex}$ for different temperatures $T_L=0.2 \mathrm{K}$, $0.25 \mathrm{K}$, and $0.3 \mathrm{K}$. We can find that heat current has non-monotonic behavior as the magnetic flux $\varphi_{ex}$ increases. Moreover, the heat current is symmetric on $\varphi_{ex}/2 \pi=0.5$. When $0\leq \varphi_{ex}/2 \pi \leq 0.5$, heat current firstly decreases and then increases as the external flux increases. One can obtain the maximal heat current when $\varphi_{ex}/2 \pi=0.5$. At this point, the transition frequencies $\omega_{\pm}$ have the minimal local value and can exchange the maximal energy for fixed temperatures of the two terminals. Thus, one can use external flux to control heat transfer, so this system can be considered a heat valve.  However, for $\varphi_{ex}/2 \pi\simeq 0.374$ and $\varphi_{ex}/2 \pi\simeq 0.624$, $\dot{\mathcal{Q}}_{L}=0$, i.e., the two resonators are entirely independent and have no energy exchange as expected. This result is similar to the optomechanical system in Ref. \cite{PhysRevA.98.052123}. In addition, we also find that the more temperature bias, the larger the heat current, which is consistent with our intuition. 
It is clear that the heat is modulated by changing the external flux  $\varphi_{ex}$  in a wide range, actually implying that the system undergoes from the weak, strong, to ultrastrong coupling regimes. This is why we emphasize the wide-range tunability of the coupling strength realized in Ref. \cite{PhysRevApplied.16.064041}.

To clarify the potential relation of quantum heat transport and entanglement of two resonators, we plot both heat current and entanglement in Fig. \ref{fig: Heat_EN}. The figure shows that the external flux can also modulate the entanglement of two coupling resonators. Fig. \ref{fig: Heat_EN} indicates that the entanglement of two coupled resonators demonstrates an entirely similar trend to the heat current. They have good consistency in the given parameter range. They all depend on the values of $a$ and $b$ as shown in Eq. (\ref{hc dd}) and the matrix elements of Eq. (\ref{Cov Matrix}).  In particular, one can analytically obtain that the two minimal values of heat current and entanglement are achieved at $g=0$, i.e.,  $M_{*} (\varphi_{\mathrm{ex}})=0$. The corresponding $\varphi_{\mathrm{ex}}$ can be solved as $\varphi_{\mathrm{ex1}}= \arccos \left(-L_{J 0}/(\frac{L_{sh}^{2}}{M_0}+2 L_{sh}+L_{0})\right)+\delta$ and $\varphi_{\mathrm{ex2}}=2 \pi- \varphi_{\mathrm{ex1}}$  due to the periodicity property.  The maximum heat current and entanglement value is at $\varphi_{\mathrm{ex}}/2 \pi=0.5$.

\begin{figure}
\includegraphics[width=97 mm]{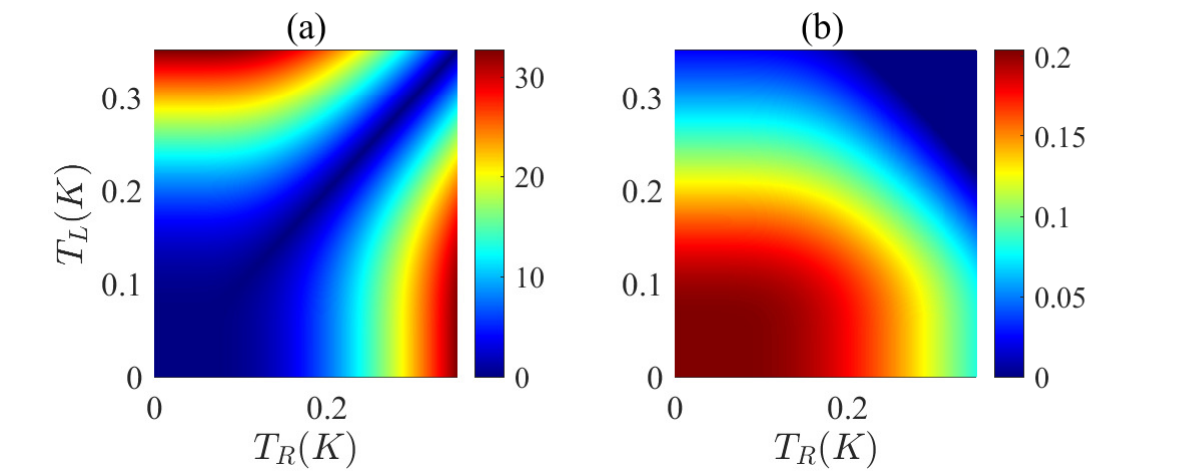}
\caption{\label{fig: Heat_EN_tem}
Heat currents (a) and entanglement (b) versus temperatures $T_R$  and $T_L$.  Here $\varphi_{ex}/2 \pi=0.5$ and other parameters are the same in Fig. \ref{fig: mod_omega_g}.}
\end{figure}

In addition, Fig. \ref{fig: Heat_EN} indicates that for $T_L>T_R$, the larger $T_L$, the smaller entanglement, contrasts the heat currents. 
To further illustrate the effect of temperature, we plot how the heat currents and entanglement depend on the temperature in Fig. \ref{fig: Heat_EN_tem}.  It is clearly shown that the heat currents depend on the temperature bias. Namely,  the larger temperature leads to a large heat current, but quantum entanglement decreases monotonically as $T_L$ or $T_R$ increases. However,  both have consistent behavior as the change of the external flux $\varphi_{ex}$.

Finally,  we briefly discuss possible experimental detection of heat transport and entanglement. It is shown that biased normal metal-insulator-superconductor junctions can be used to monitor and control temperature \cite{RevModPhys.78.217}, and heat current can be measured by the standard normal-metal-insulator-superconductor (NIS) thermometry techniques \cite{PhysRevApplied.17.064022}. The more detailed discussions concerning the experimental scenario can refer to Refs. \cite{senior2020heat, ronzani2018tunable}. The entanglement is completely characterized by the covariance matrix so that it can be detected by the covariance matrix tomography similar to Refs. \cite{barzanjeh2019stationary,chen2020entanglement,kotler2021direct}.

\label{conclusion}
In conclusion, we have designed a flux-tunable heat valve through ultrastrong $L C$ resonators, which can be modulated via the magnetic flux employing a SQUID.  This is experimentally realizable due to the current quantum superconducting technology.
We find that quantum entanglement in the system shows consistent behavior compared to heat transfer, which implies that quantum entanglement plays a dominant role in the heat valve. This work can provide an insightful reference to understand the quantum features in quantum thermal machines.

\bigskip
See the supplementary material for diagonalization of Hamiltonian, derivation of master equation, and steady state solution. 

\bigskip
This work was supported by the National Natural Science Foundation of China under Grants No.12175029, No.
12011530014 and No.11775040, and the Key Research and
Development Project of Liaoning Province, under Grant No.
2020JH2/10500003.

\bigskip
The data that support the findings of this study are available from
the corresponding author upon reasonable request.

\section*{References}
\bibliography{ref}

\providecommand{\noopsort}[1]{}\providecommand{\singleletter}[1]{#1}%
\begin{thebibliography}{62}%
\makeatletter
\providecommand \@ifxundefined [1]{%
 \@ifx{#1\undefined}
}%
\providecommand \@ifnum [1]{%
 \ifnum #1\expandafter \@firstoftwo
 \else \expandafter \@secondoftwo
 \fi
}%
\providecommand \@ifx [1]{%
 \ifx #1\expandafter \@firstoftwo
 \else \expandafter \@secondoftwo
 \fi
}%
\providecommand \natexlab [1]{#1}%
\providecommand \enquote  [1]{``#1''}%
\providecommand \bibnamefont  [1]{#1}%
\providecommand \bibfnamefont [1]{#1}%
\providecommand \citenamefont [1]{#1}%
\providecommand \href@noop [0]{\@secondoftwo}%
\providecommand \href [0]{\begingroup \@sanitize@url \@href}%
\providecommand \@href[1]{\@@startlink{#1}\@@href}%
\providecommand \@@href[1]{\endgroup#1\@@endlink}%
\providecommand \@sanitize@url [0]{\catcode `\\12\catcode `\$12\catcode
  `\&12\catcode `\#12\catcode `\^12\catcode `\_12\catcode `\%12\relax}%
\providecommand \@@startlink[1]{}%
\providecommand \@@endlink[0]{}%
\providecommand \url  [0]{\begingroup\@sanitize@url \@url }%
\providecommand \@url [1]{\endgroup\@href {#1}{\urlprefix }}%
\providecommand \urlprefix  [0]{URL }%
\providecommand \Eprint [0]{\href }%
\providecommand \doibase [0]{https://doi.org/}%
\providecommand \selectlanguage [0]{\@gobble}%
\providecommand \bibinfo  [0]{\@secondoftwo}%
\providecommand \bibfield  [0]{\@secondoftwo}%
\providecommand \translation [1]{[#1]}%
\providecommand \BibitemOpen [0]{}%
\providecommand \bibitemStop [0]{}%
\providecommand \bibitemNoStop [0]{.\EOS\space}%
\providecommand \EOS [0]{\spacefactor3000\relax}%
\providecommand \BibitemShut  [1]{\csname bibitem#1\endcsname}%
\let\auto@bib@innerbib\@empty
\bibitem [{\citenamefont {Vinjanampathy}\ and\ \citenamefont
  {Anders}(2016)}]{vinjanampathy2016quantum}%
  \BibitemOpen
  \bibfield  {author} {\bibinfo {author} {\bibfnamefont {S.}~\bibnamefont
  {Vinjanampathy}}\ and\ \bibinfo {author} {\bibfnamefont {J.}~\bibnamefont
  {Anders}},\ }\bibfield  {title} {\enquote {\bibinfo {title} {Quantum
  thermodynamics},}\ }\href@noop {} {\bibfield  {journal} {\bibinfo  {journal}
  {Contemp. Phys.}\ }\textbf {\bibinfo {volume} {57}},\ \bibinfo {pages}
  {545--579} (\bibinfo {year} {2016})}\BibitemShut {NoStop}%
\bibitem [{\citenamefont {Pekola}\ and\ \citenamefont
  {Karimi}(2021)}]{RevModPhys.93.041001}%
  \BibitemOpen
  \bibfield  {author} {\bibinfo {author} {\bibfnamefont {J.~P.}\ \bibnamefont
  {Pekola}}\ and\ \bibinfo {author} {\bibfnamefont {B.}~\bibnamefont
  {Karimi}},\ }\bibfield  {title} {\enquote {\bibinfo {title} {Colloquium:
  Quantum heat transport in condensed matter systems},}\ }\href
  {https://doi.org/10.1103/RevModPhys.93.041001} {\bibfield  {journal}
  {\bibinfo  {journal} {Rev. Mod. Phys.}\ }\textbf {\bibinfo {volume} {93}},\
  \bibinfo {pages} {041001} (\bibinfo {year} {2021})}\BibitemShut {NoStop}%
\bibitem [{\citenamefont {Joulain}\ \emph {et~al.}(2016)\citenamefont
  {Joulain}, \citenamefont {Drevillon}, \citenamefont {Ezzahri},\ and\
  \citenamefont {Ordonez-Miranda}}]{PhysRevLett.116.200601}%
  \BibitemOpen
  \bibfield  {author} {\bibinfo {author} {\bibfnamefont {K.}~\bibnamefont
  {Joulain}}, \bibinfo {author} {\bibfnamefont {J.}~\bibnamefont {Drevillon}},
  \bibinfo {author} {\bibfnamefont {Y.}~\bibnamefont {Ezzahri}},\ and\ \bibinfo
  {author} {\bibfnamefont {J.}~\bibnamefont {Ordonez-Miranda}},\ }\bibfield
  {title} {\enquote {\bibinfo {title} {Quantum thermal transistor},}\ }\href
  {https://doi.org/10.1103/PhysRevLett.116.200601} {\bibfield  {journal}
  {\bibinfo  {journal} {Phys. Rev. Lett.}\ }\textbf {\bibinfo {volume} {116}},\
  \bibinfo {pages} {200601} (\bibinfo {year} {2016})}\BibitemShut {NoStop}%
\bibitem [{\citenamefont {Werlang}\ \emph {et~al.}(2014)\citenamefont
  {Werlang}, \citenamefont {Marchiori}, \citenamefont {Cornelio},\ and\
  \citenamefont {Valente}}]{PhysRevE.89.062109}%
  \BibitemOpen
  \bibfield  {author} {\bibinfo {author} {\bibfnamefont {T.}~\bibnamefont
  {Werlang}}, \bibinfo {author} {\bibfnamefont {M.~A.}\ \bibnamefont
  {Marchiori}}, \bibinfo {author} {\bibfnamefont {M.~F.}\ \bibnamefont
  {Cornelio}},\ and\ \bibinfo {author} {\bibfnamefont {D.}~\bibnamefont
  {Valente}},\ }\bibfield  {title} {\enquote {\bibinfo {title} {Optimal
  rectification in the ultrastrong coupling regime},}\ }\href
  {https://doi.org/10.1103/PhysRevE.89.062109} {\bibfield  {journal} {\bibinfo
  {journal} {Phys. Rev. E}\ }\textbf {\bibinfo {volume} {89}},\ \bibinfo
  {pages} {062109} (\bibinfo {year} {2014})}\BibitemShut {NoStop}%
\bibitem [{\citenamefont {Ordonez-Miranda}, \citenamefont {Ezzahri},\ and\
  \citenamefont {Joulain}(2017)}]{PhysRevE.95.022128}%
  \BibitemOpen
  \bibfield  {author} {\bibinfo {author} {\bibfnamefont {J.}~\bibnamefont
  {Ordonez-Miranda}}, \bibinfo {author} {\bibfnamefont {Y.}~\bibnamefont
  {Ezzahri}},\ and\ \bibinfo {author} {\bibfnamefont {K.}~\bibnamefont
  {Joulain}},\ }\bibfield  {title} {\enquote {\bibinfo {title} {Quantum thermal
  diode based on two interacting spinlike systems under different
  excitations},}\ }\href {https://doi.org/10.1103/PhysRevE.95.022128}
  {\bibfield  {journal} {\bibinfo  {journal} {Phys. Rev. E}\ }\textbf {\bibinfo
  {volume} {95}},\ \bibinfo {pages} {022128} (\bibinfo {year}
  {2017})}\BibitemShut {NoStop}%
\bibitem [{\citenamefont {Karg\ifmmode \imath \else~\i \fi{}}\ \emph
  {et~al.}(2019)\citenamefont {Karg\ifmmode \imath \else~\i \fi{}},
  \citenamefont {Naseem}, \citenamefont {Opatrn\'y}, \citenamefont
  {M\"ustecapl\ifmmode \imath \else \i \fi{}o\ifmmode~\breve{g}\else
  \u{g}\fi{}lu},\ and\ \citenamefont {Kurizki}}]{PhysRevE.99.042121}%
  \BibitemOpen
  \bibfield  {author} {\bibinfo {author} {\bibfnamefont {C.}~\bibnamefont
  {Karg\ifmmode \imath \else~\i \fi{}}}, \bibinfo {author} {\bibfnamefont
  {M.~T.}\ \bibnamefont {Naseem}}, \bibinfo {author} {\bibfnamefont {T.~c.~v.}\
  \bibnamefont {Opatrn\'y}}, \bibinfo {author} {\bibfnamefont {O.~E.}\
  \bibnamefont {M\"ustecapl\ifmmode \imath \else \i
  \fi{}o\ifmmode~\breve{g}\else \u{g}\fi{}lu}},\ and\ \bibinfo {author}
  {\bibfnamefont {G.}~\bibnamefont {Kurizki}},\ }\bibfield  {title} {\enquote
  {\bibinfo {title} {Quantum optical two-atom thermal diode},}\ }\href
  {https://doi.org/10.1103/PhysRevE.99.042121} {\bibfield  {journal} {\bibinfo
  {journal} {Phys. Rev. E}\ }\textbf {\bibinfo {volume} {99}},\ \bibinfo
  {pages} {042121} (\bibinfo {year} {2019})}\BibitemShut {NoStop}%
\bibitem [{\citenamefont {Guo}, \citenamefont {Liu},\ and\ \citenamefont
  {Yu}(2019)}]{PhysRevE.99.032112}%
  \BibitemOpen
  \bibfield  {author} {\bibinfo {author} {\bibfnamefont {B.-q.}\ \bibnamefont
  {Guo}}, \bibinfo {author} {\bibfnamefont {T.}~\bibnamefont {Liu}},\ and\
  \bibinfo {author} {\bibfnamefont {C.-s.}\ \bibnamefont {Yu}},\ }\bibfield
  {title} {\enquote {\bibinfo {title} {Multifunctional quantum thermal device
  utilizing three qubits},}\ }\href
  {https://doi.org/10.1103/PhysRevE.99.032112} {\bibfield  {journal} {\bibinfo
  {journal} {Phys. Rev. E}\ }\textbf {\bibinfo {volume} {99}},\ \bibinfo
  {pages} {032112} (\bibinfo {year} {2019})}\BibitemShut {NoStop}%
\bibitem [{\citenamefont {Gupt}\ \emph {et~al.}(2022)\citenamefont {Gupt},
  \citenamefont {Bhattacharyya}, \citenamefont {Das}, \citenamefont {Datta},
  \citenamefont {Mukherjee},\ and\ \citenamefont
  {Ghosh}}]{PhysRevE.106.024110}%
  \BibitemOpen
  \bibfield  {author} {\bibinfo {author} {\bibfnamefont {N.}~\bibnamefont
  {Gupt}}, \bibinfo {author} {\bibfnamefont {S.}~\bibnamefont {Bhattacharyya}},
  \bibinfo {author} {\bibfnamefont {B.}~\bibnamefont {Das}}, \bibinfo {author}
  {\bibfnamefont {S.}~\bibnamefont {Datta}}, \bibinfo {author} {\bibfnamefont
  {V.}~\bibnamefont {Mukherjee}},\ and\ \bibinfo {author} {\bibfnamefont
  {A.}~\bibnamefont {Ghosh}},\ }\bibfield  {title} {\enquote {\bibinfo {title}
  {Floquet quantum thermal transistor},}\ }\href
  {https://doi.org/10.1103/PhysRevE.106.024110} {\bibfield  {journal} {\bibinfo
   {journal} {Phys. Rev. E}\ }\textbf {\bibinfo {volume} {106}},\ \bibinfo
  {pages} {024110} (\bibinfo {year} {2022})}\BibitemShut {NoStop}%
\bibitem [{\citenamefont {Aligia}\ \emph {et~al.}(2020)\citenamefont {Aligia},
  \citenamefont {Daroca}, \citenamefont {Arrachea},\ and\ \citenamefont
  {Roura-Bas}}]{PhysRevB.101.075417}%
  \BibitemOpen
  \bibfield  {author} {\bibinfo {author} {\bibfnamefont {A.~A.}\ \bibnamefont
  {Aligia}}, \bibinfo {author} {\bibfnamefont {D.~P.}\ \bibnamefont {Daroca}},
  \bibinfo {author} {\bibfnamefont {L.}~\bibnamefont {Arrachea}},\ and\
  \bibinfo {author} {\bibfnamefont {P.}~\bibnamefont {Roura-Bas}},\ }\bibfield
  {title} {\enquote {\bibinfo {title} {Heat current across a capacitively
  coupled double quantum dot},}\ }\href
  {https://doi.org/10.1103/PhysRevB.101.075417} {\bibfield  {journal} {\bibinfo
   {journal} {Phys. Rev. B}\ }\textbf {\bibinfo {volume} {101}},\ \bibinfo
  {pages} {075417} (\bibinfo {year} {2020})}\BibitemShut {NoStop}%
\bibitem [{\citenamefont {Levy}, \citenamefont {Alicki},\ and\ \citenamefont
  {Kosloff}(2012)}]{PhysRevE.85.061126}%
  \BibitemOpen
  \bibfield  {author} {\bibinfo {author} {\bibfnamefont {A.}~\bibnamefont
  {Levy}}, \bibinfo {author} {\bibfnamefont {R.}~\bibnamefont {Alicki}},\ and\
  \bibinfo {author} {\bibfnamefont {R.}~\bibnamefont {Kosloff}},\ }\bibfield
  {title} {\enquote {\bibinfo {title} {Quantum refrigerators and the third law
  of thermodynamics},}\ }\href {https://doi.org/10.1103/PhysRevE.85.061126}
  {\bibfield  {journal} {\bibinfo  {journal} {Phys. Rev. E}\ }\textbf {\bibinfo
  {volume} {85}},\ \bibinfo {pages} {061126} (\bibinfo {year}
  {2012})}\BibitemShut {NoStop}%
\bibitem [{\citenamefont {Hofer}\ \emph {et~al.}(2016)\citenamefont {Hofer},
  \citenamefont {Perarnau-Llobet}, \citenamefont {Brask}, \citenamefont
  {Silva}, \citenamefont {Huber},\ and\ \citenamefont
  {Brunner}}]{PhysRevB.94.235420}%
  \BibitemOpen
  \bibfield  {author} {\bibinfo {author} {\bibfnamefont {P.~P.}\ \bibnamefont
  {Hofer}}, \bibinfo {author} {\bibfnamefont {M.}~\bibnamefont
  {Perarnau-Llobet}}, \bibinfo {author} {\bibfnamefont {J.~B.}\ \bibnamefont
  {Brask}}, \bibinfo {author} {\bibfnamefont {R.}~\bibnamefont {Silva}},
  \bibinfo {author} {\bibfnamefont {M.}~\bibnamefont {Huber}},\ and\ \bibinfo
  {author} {\bibfnamefont {N.}~\bibnamefont {Brunner}},\ }\bibfield  {title}
  {\enquote {\bibinfo {title} {{Autonomous quantum refrigerator in a circuit
  QED architecture based on a Josephson junction}},}\ }\href
  {https://doi.org/10.1103/PhysRevB.94.235420} {\bibfield  {journal} {\bibinfo
  {journal} {Phys. Rev. B}\ }\textbf {\bibinfo {volume} {94}},\ \bibinfo
  {pages} {235420} (\bibinfo {year} {2016})}\BibitemShut {NoStop}%
\bibitem [{\citenamefont {Tan}\ \emph {et~al.}(2017)\citenamefont {Tan},
  \citenamefont {Partanen}, \citenamefont {Lake}, \citenamefont {Govenius},
  \citenamefont {Masuda},\ and\ \citenamefont
  {M{\"o}tt{\"o}nen}}]{tan2017quantum}%
  \BibitemOpen
  \bibfield  {author} {\bibinfo {author} {\bibfnamefont {K.~Y.}\ \bibnamefont
  {Tan}}, \bibinfo {author} {\bibfnamefont {M.}~\bibnamefont {Partanen}},
  \bibinfo {author} {\bibfnamefont {R.~E.}\ \bibnamefont {Lake}}, \bibinfo
  {author} {\bibfnamefont {J.}~\bibnamefont {Govenius}}, \bibinfo {author}
  {\bibfnamefont {S.}~\bibnamefont {Masuda}},\ and\ \bibinfo {author}
  {\bibfnamefont {M.}~\bibnamefont {M{\"o}tt{\"o}nen}},\ }\bibfield  {title}
  {\enquote {\bibinfo {title} {Quantum-circuit refrigerator},}\ }\href@noop {}
  {\bibfield  {journal} {\bibinfo  {journal} {Nat. Commun.}\ }\textbf {\bibinfo
  {volume} {8}},\ \bibinfo {pages} {1--8} (\bibinfo {year} {2017})}\BibitemShut
  {NoStop}%
\bibitem [{\citenamefont {Gubaydullin}\ \emph {et~al.}(2022)\citenamefont
  {Gubaydullin}, \citenamefont {Thomas}, \citenamefont {Golubev}, \citenamefont
  {Lvov}, \citenamefont {Peltonen},\ and\ \citenamefont
  {Pekola}}]{gubaydullin2022photonic}%
  \BibitemOpen
  \bibfield  {author} {\bibinfo {author} {\bibfnamefont {A.}~\bibnamefont
  {Gubaydullin}}, \bibinfo {author} {\bibfnamefont {G.}~\bibnamefont {Thomas}},
  \bibinfo {author} {\bibfnamefont {D.~S.}\ \bibnamefont {Golubev}}, \bibinfo
  {author} {\bibfnamefont {D.}~\bibnamefont {Lvov}}, \bibinfo {author}
  {\bibfnamefont {J.~T.}\ \bibnamefont {Peltonen}},\ and\ \bibinfo {author}
  {\bibfnamefont {J.~P.}\ \bibnamefont {Pekola}},\ }\bibfield  {title}
  {\enquote {\bibinfo {title} {Photonic heat transport in three terminal
  superconducting circuit},}\ }\href@noop {} {\bibfield  {journal} {\bibinfo
  {journal} {Nat. Commun.}\ }\textbf {\bibinfo {volume} {13}},\ \bibinfo
  {pages} {1552} (\bibinfo {year} {2022})}\BibitemShut {NoStop}%
\bibitem [{\citenamefont {Elouard}\ \emph {et~al.}(2020)\citenamefont
  {Elouard}, \citenamefont {Thomas}, \citenamefont {Maillet}, \citenamefont
  {Pekola},\ and\ \citenamefont {Jordan}}]{PhysRevE.102.030102}%
  \BibitemOpen
  \bibfield  {author} {\bibinfo {author} {\bibfnamefont {C.}~\bibnamefont
  {Elouard}}, \bibinfo {author} {\bibfnamefont {G.}~\bibnamefont {Thomas}},
  \bibinfo {author} {\bibfnamefont {O.}~\bibnamefont {Maillet}}, \bibinfo
  {author} {\bibfnamefont {J.~P.}\ \bibnamefont {Pekola}},\ and\ \bibinfo
  {author} {\bibfnamefont {A.~N.}\ \bibnamefont {Jordan}},\ }\bibfield  {title}
  {\enquote {\bibinfo {title} {Quantifying the quantum heat contribution from a
  driven superconducting circuit},}\ }\href
  {https://doi.org/10.1103/PhysRevE.102.030102} {\bibfield  {journal} {\bibinfo
   {journal} {Phys. Rev. E}\ }\textbf {\bibinfo {volume} {102}},\ \bibinfo
  {pages} {030102} (\bibinfo {year} {2020})}\BibitemShut {NoStop}%
\bibitem [{\citenamefont {Hofer}\ \emph {et~al.}(2017)\citenamefont {Hofer},
  \citenamefont {Brask}, \citenamefont {Perarnau-Llobet},\ and\ \citenamefont
  {Brunner}}]{PhysRevLett.119.090603}%
  \BibitemOpen
  \bibfield  {author} {\bibinfo {author} {\bibfnamefont {P.~P.}\ \bibnamefont
  {Hofer}}, \bibinfo {author} {\bibfnamefont {J.~B.}\ \bibnamefont {Brask}},
  \bibinfo {author} {\bibfnamefont {M.}~\bibnamefont {Perarnau-Llobet}},\ and\
  \bibinfo {author} {\bibfnamefont {N.}~\bibnamefont {Brunner}},\ }\bibfield
  {title} {\enquote {\bibinfo {title} {Quantum thermal machine as a
  thermometer},}\ }\href {https://doi.org/10.1103/PhysRevLett.119.090603}
  {\bibfield  {journal} {\bibinfo  {journal} {Phys. Rev. Lett.}\ }\textbf
  {\bibinfo {volume} {119}},\ \bibinfo {pages} {090603} (\bibinfo {year}
  {2017})}\BibitemShut {NoStop}%
\bibitem [{\citenamefont {Liu}, \citenamefont {Yang},\ and\ \citenamefont
  {Yu}(2023)}]{PhysRevE.107.044121}%
  \BibitemOpen
  \bibfield  {author} {\bibinfo {author} {\bibfnamefont {Y.-q.}\ \bibnamefont
  {Liu}}, \bibinfo {author} {\bibfnamefont {Y.-j.}\ \bibnamefont {Yang}},\ and\
  \bibinfo {author} {\bibfnamefont {C.-s.}\ \bibnamefont {Yu}},\ }\bibfield
  {title} {\enquote {\bibinfo {title} {Quantum heat diode versus light emission
  in circuit quantum electrodynamical system},}\ }\href
  {https://doi.org/10.1103/PhysRevE.107.044121} {\bibfield  {journal} {\bibinfo
   {journal} {Phys. Rev. E}\ }\textbf {\bibinfo {volume} {107}},\ \bibinfo
  {pages} {044121} (\bibinfo {year} {2023})}\BibitemShut {NoStop}%
\bibitem [{\citenamefont {Forn-D\'{\i}az}\ \emph {et~al.}(2019)\citenamefont
  {Forn-D\'{\i}az}, \citenamefont {Lamata}, \citenamefont {Rico}, \citenamefont
  {Kono},\ and\ \citenamefont {Solano}}]{RevModPhys.91.025005}%
  \BibitemOpen
  \bibfield  {author} {\bibinfo {author} {\bibfnamefont {P.}~\bibnamefont
  {Forn-D\'{\i}az}}, \bibinfo {author} {\bibfnamefont {L.}~\bibnamefont
  {Lamata}}, \bibinfo {author} {\bibfnamefont {E.}~\bibnamefont {Rico}},
  \bibinfo {author} {\bibfnamefont {J.}~\bibnamefont {Kono}},\ and\ \bibinfo
  {author} {\bibfnamefont {E.}~\bibnamefont {Solano}},\ }\bibfield  {title}
  {\enquote {\bibinfo {title} {Ultrastrong coupling regimes of light-matter
  interaction},}\ }\href {https://doi.org/10.1103/RevModPhys.91.025005}
  {\bibfield  {journal} {\bibinfo  {journal} {Rev. Mod. Phys.}\ }\textbf
  {\bibinfo {volume} {91}},\ \bibinfo {pages} {025005} (\bibinfo {year}
  {2019})}\BibitemShut {NoStop}%
\bibitem [{\citenamefont {Blais}\ \emph {et~al.}(2021)\citenamefont {Blais},
  \citenamefont {Grimsmo}, \citenamefont {Girvin},\ and\ \citenamefont
  {Wallraff}}]{RevModPhys.93.025005}%
  \BibitemOpen
  \bibfield  {author} {\bibinfo {author} {\bibfnamefont {A.}~\bibnamefont
  {Blais}}, \bibinfo {author} {\bibfnamefont {A.~L.}\ \bibnamefont {Grimsmo}},
  \bibinfo {author} {\bibfnamefont {S.~M.}\ \bibnamefont {Girvin}},\ and\
  \bibinfo {author} {\bibfnamefont {A.}~\bibnamefont {Wallraff}},\ }\bibfield
  {title} {\enquote {\bibinfo {title} {Circuit quantum electrodynamics},}\
  }\href {https://doi.org/10.1103/RevModPhys.93.025005} {\bibfield  {journal}
  {\bibinfo  {journal} {Rev. Mod. Phys.}\ }\textbf {\bibinfo {volume} {93}},\
  \bibinfo {pages} {025005} (\bibinfo {year} {2021})}\BibitemShut {NoStop}%
\bibitem [{\citenamefont {Felicetti}, \citenamefont {Hwang},\ and\
  \citenamefont {Le~Boit\'e}(2018)}]{PhysRevA.98.053859}%
  \BibitemOpen
  \bibfield  {author} {\bibinfo {author} {\bibfnamefont {S.}~\bibnamefont
  {Felicetti}}, \bibinfo {author} {\bibfnamefont {M.-J.}\ \bibnamefont
  {Hwang}},\ and\ \bibinfo {author} {\bibfnamefont {A.}~\bibnamefont
  {Le~Boit\'e}},\ }\bibfield  {title} {\enquote {\bibinfo {title}
  {Ultrastrong-coupling regime of nondipolar light-matter interactions},}\
  }\href {https://doi.org/10.1103/PhysRevA.98.053859} {\bibfield  {journal}
  {\bibinfo  {journal} {Phys. Rev. A}\ }\textbf {\bibinfo {volume} {98}},\
  \bibinfo {pages} {053859} (\bibinfo {year} {2018})}\BibitemShut {NoStop}%
\bibitem [{\citenamefont {Yoshihara}\ \emph {et~al.}(2017)\citenamefont
  {Yoshihara}, \citenamefont {Fuse}, \citenamefont {Ashhab}, \citenamefont
  {Kakuyanagi}, \citenamefont {Saito},\ and\ \citenamefont
  {Semba}}]{PhysRevA.95.053824}%
  \BibitemOpen
  \bibfield  {author} {\bibinfo {author} {\bibfnamefont {F.}~\bibnamefont
  {Yoshihara}}, \bibinfo {author} {\bibfnamefont {T.}~\bibnamefont {Fuse}},
  \bibinfo {author} {\bibfnamefont {S.}~\bibnamefont {Ashhab}}, \bibinfo
  {author} {\bibfnamefont {K.}~\bibnamefont {Kakuyanagi}}, \bibinfo {author}
  {\bibfnamefont {S.}~\bibnamefont {Saito}},\ and\ \bibinfo {author}
  {\bibfnamefont {K.}~\bibnamefont {Semba}},\ }\bibfield  {title} {\enquote
  {\bibinfo {title} {Characteristic spectra of circuit quantum electrodynamics
  systems from the ultrastrong- to the deep-strong-coupling regime},}\ }\href
  {https://doi.org/10.1103/PhysRevA.95.053824} {\bibfield  {journal} {\bibinfo
  {journal} {Phys. Rev. A}\ }\textbf {\bibinfo {volume} {95}},\ \bibinfo
  {pages} {053824} (\bibinfo {year} {2017})}\BibitemShut {NoStop}%
\bibitem [{\citenamefont {Beaudoin}, \citenamefont {Gambetta},\ and\
  \citenamefont {Blais}(2011)}]{PhysRevA.84.043832}%
  \BibitemOpen
  \bibfield  {author} {\bibinfo {author} {\bibfnamefont {F.}~\bibnamefont
  {Beaudoin}}, \bibinfo {author} {\bibfnamefont {J.~M.}\ \bibnamefont
  {Gambetta}},\ and\ \bibinfo {author} {\bibfnamefont {A.}~\bibnamefont
  {Blais}},\ }\bibfield  {title} {\enquote {\bibinfo {title} {{Dissipation and
  ultrastrong coupling in circuit QED}},}\ }\href
  {https://doi.org/10.1103/PhysRevA.84.043832} {\bibfield  {journal} {\bibinfo
  {journal} {Phys. Rev. A}\ }\textbf {\bibinfo {volume} {84}},\ \bibinfo
  {pages} {043832} (\bibinfo {year} {2011})}\BibitemShut {NoStop}%
\bibitem [{\citenamefont {Frisk~Kockum}\ \emph {et~al.}(2019)\citenamefont
  {Frisk~Kockum}, \citenamefont {Miranowicz}, \citenamefont {De~Liberato},
  \citenamefont {Savasta},\ and\ \citenamefont {Nori}}]{frisk2019ultrastrong}%
  \BibitemOpen
  \bibfield  {author} {\bibinfo {author} {\bibfnamefont {A.}~\bibnamefont
  {Frisk~Kockum}}, \bibinfo {author} {\bibfnamefont {A.}~\bibnamefont
  {Miranowicz}}, \bibinfo {author} {\bibfnamefont {S.}~\bibnamefont
  {De~Liberato}}, \bibinfo {author} {\bibfnamefont {S.}~\bibnamefont
  {Savasta}},\ and\ \bibinfo {author} {\bibfnamefont {F.}~\bibnamefont
  {Nori}},\ }\bibfield  {title} {\enquote {\bibinfo {title} {Ultrastrong
  coupling between light and matter},}\ }\href@noop {} {\bibfield  {journal}
  {\bibinfo  {journal} {Nat. Rev. Phys.}\ }\textbf {\bibinfo {volume} {1}},\
  \bibinfo {pages} {19--40} (\bibinfo {year} {2019})}\BibitemShut {NoStop}%
\bibitem [{\citenamefont {Upadhyay}\ \emph {et~al.}(2021)\citenamefont
  {Upadhyay}, \citenamefont {Thomas}, \citenamefont {Chang}, \citenamefont
  {Golubev}, \citenamefont {Guthrie}, \citenamefont {Gubaydullin},
  \citenamefont {Peltonen},\ and\ \citenamefont
  {Pekola}}]{PhysRevApplied.16.044045}%
  \BibitemOpen
  \bibfield  {author} {\bibinfo {author} {\bibfnamefont {R.}~\bibnamefont
  {Upadhyay}}, \bibinfo {author} {\bibfnamefont {G.}~\bibnamefont {Thomas}},
  \bibinfo {author} {\bibfnamefont {Y.-C.}\ \bibnamefont {Chang}}, \bibinfo
  {author} {\bibfnamefont {D.~S.}\ \bibnamefont {Golubev}}, \bibinfo {author}
  {\bibfnamefont {A.}~\bibnamefont {Guthrie}}, \bibinfo {author} {\bibfnamefont
  {A.}~\bibnamefont {Gubaydullin}}, \bibinfo {author} {\bibfnamefont {J.~T.}\
  \bibnamefont {Peltonen}},\ and\ \bibinfo {author} {\bibfnamefont {J.~P.}\
  \bibnamefont {Pekola}},\ }\bibfield  {title} {\enquote {\bibinfo {title}
  {Robust strong-coupling architecture in circuit quantum electrodynamics},}\
  }\href {https://doi.org/10.1103/PhysRevApplied.16.044045} {\bibfield
  {journal} {\bibinfo  {journal} {Phys. Rev. Appl.}\ }\textbf {\bibinfo
  {volume} {16}},\ \bibinfo {pages} {044045} (\bibinfo {year}
  {2021})}\BibitemShut {NoStop}%
\bibitem [{\citenamefont {Niemczyk}\ \emph {et~al.}(2010)\citenamefont
  {Niemczyk}, \citenamefont {Deppe}, \citenamefont {Huebl}, \citenamefont
  {Menzel}, \citenamefont {Hocke}, \citenamefont {Schwarz}, \citenamefont
  {Garcia-Ripoll}, \citenamefont {Zueco}, \citenamefont {H{\"u}mmer},
  \citenamefont {Solano} \emph {et~al.}}]{niemczyk2010circuit}%
  \BibitemOpen
  \bibfield  {author} {\bibinfo {author} {\bibfnamefont {T.}~\bibnamefont
  {Niemczyk}}, \bibinfo {author} {\bibfnamefont {F.}~\bibnamefont {Deppe}},
  \bibinfo {author} {\bibfnamefont {H.}~\bibnamefont {Huebl}}, \bibinfo
  {author} {\bibfnamefont {E.}~\bibnamefont {Menzel}}, \bibinfo {author}
  {\bibfnamefont {F.}~\bibnamefont {Hocke}}, \bibinfo {author} {\bibfnamefont
  {M.}~\bibnamefont {Schwarz}}, \bibinfo {author} {\bibfnamefont
  {J.}~\bibnamefont {Garcia-Ripoll}}, \bibinfo {author} {\bibfnamefont
  {D.}~\bibnamefont {Zueco}}, \bibinfo {author} {\bibfnamefont
  {T.}~\bibnamefont {H{\"u}mmer}}, \bibinfo {author} {\bibfnamefont
  {E.}~\bibnamefont {Solano}}, \emph {et~al.},\ }\bibfield  {title} {\enquote
  {\bibinfo {title} {Circuit quantum electrodynamics in the
  ultrastrong-coupling regime},}\ }\href@noop {} {\bibfield  {journal}
  {\bibinfo  {journal} {Nat. Phys.}\ }\textbf {\bibinfo {volume} {6}},\
  \bibinfo {pages} {772--776} (\bibinfo {year} {2010})}\BibitemShut {NoStop}%
\bibitem [{\citenamefont {Karimi}\ and\ \citenamefont
  {Pekola}(2017)}]{PhysRevB.96.115408}%
  \BibitemOpen
  \bibfield  {author} {\bibinfo {author} {\bibfnamefont {B.}~\bibnamefont
  {Karimi}}\ and\ \bibinfo {author} {\bibfnamefont {J.~P.}\ \bibnamefont
  {Pekola}},\ }\bibfield  {title} {\enquote {\bibinfo {title} {Correlated
  versus uncorrelated noise acting on a quantum refrigerator},}\ }\href
  {https://doi.org/10.1103/PhysRevB.96.115408} {\bibfield  {journal} {\bibinfo
  {journal} {Phys. Rev. B}\ }\textbf {\bibinfo {volume} {96}},\ \bibinfo
  {pages} {115408} (\bibinfo {year} {2017})}\BibitemShut {NoStop}%
\bibitem [{\citenamefont {Ronzani}\ \emph {et~al.}(2018)\citenamefont
  {Ronzani}, \citenamefont {Karimi}, \citenamefont {Senior}, \citenamefont
  {Chang}, \citenamefont {Peltonen}, \citenamefont {Chen},\ and\ \citenamefont
  {Pekola}}]{ronzani2018tunable}%
  \BibitemOpen
  \bibfield  {author} {\bibinfo {author} {\bibfnamefont {A.}~\bibnamefont
  {Ronzani}}, \bibinfo {author} {\bibfnamefont {B.}~\bibnamefont {Karimi}},
  \bibinfo {author} {\bibfnamefont {J.}~\bibnamefont {Senior}}, \bibinfo
  {author} {\bibfnamefont {Y.-C.}\ \bibnamefont {Chang}}, \bibinfo {author}
  {\bibfnamefont {J.~T.}\ \bibnamefont {Peltonen}}, \bibinfo {author}
  {\bibfnamefont {C.}~\bibnamefont {Chen}},\ and\ \bibinfo {author}
  {\bibfnamefont {J.~P.}\ \bibnamefont {Pekola}},\ }\bibfield  {title}
  {\enquote {\bibinfo {title} {Tunable photonic heat transport in a quantum
  heat valve},}\ }\href@noop {} {\bibfield  {journal} {\bibinfo  {journal}
  {Nat. Phys.}\ }\textbf {\bibinfo {volume} {14}},\ \bibinfo {pages} {991--995}
  (\bibinfo {year} {2018})}\BibitemShut {NoStop}%
\bibitem [{\citenamefont {Iorio}\ \emph {et~al.}(2021)\citenamefont {Iorio},
  \citenamefont {Strambini}, \citenamefont {Haack}, \citenamefont {Campisi},\
  and\ \citenamefont {Giazotto}}]{PhysRevApplied.15.054050}%
  \BibitemOpen
  \bibfield  {author} {\bibinfo {author} {\bibfnamefont {A.}~\bibnamefont
  {Iorio}}, \bibinfo {author} {\bibfnamefont {E.}~\bibnamefont {Strambini}},
  \bibinfo {author} {\bibfnamefont {G.}~\bibnamefont {Haack}}, \bibinfo
  {author} {\bibfnamefont {M.}~\bibnamefont {Campisi}},\ and\ \bibinfo {author}
  {\bibfnamefont {F.}~\bibnamefont {Giazotto}},\ }\bibfield  {title} {\enquote
  {\bibinfo {title} {Photonic heat rectification in a system of coupled
  qubits},}\ }\href {https://doi.org/10.1103/PhysRevApplied.15.054050}
  {\bibfield  {journal} {\bibinfo  {journal} {Phys. Rev. Appl.}\ }\textbf
  {\bibinfo {volume} {15}},\ \bibinfo {pages} {054050} (\bibinfo {year}
  {2021})}\BibitemShut {NoStop}%
\bibitem [{\citenamefont {Karimi}\ \emph {et~al.}(2017)\citenamefont {Karimi},
  \citenamefont {Pekola}, \citenamefont {Campisi},\ and\ \citenamefont
  {Fazio}}]{Karimi_2017}%
  \BibitemOpen
  \bibfield  {author} {\bibinfo {author} {\bibfnamefont {B.}~\bibnamefont
  {Karimi}}, \bibinfo {author} {\bibfnamefont {J.~P.}\ \bibnamefont {Pekola}},
  \bibinfo {author} {\bibfnamefont {M.}~\bibnamefont {Campisi}},\ and\ \bibinfo
  {author} {\bibfnamefont {R.}~\bibnamefont {Fazio}},\ }\bibfield  {title}
  {\enquote {\bibinfo {title} {Coupled qubits as a quantum heat switch},}\
  }\href {https://doi.org/10.1088/2058-9565/aa8330} {\bibfield  {journal}
  {\bibinfo  {journal} {Quantum Science and Technology}\ }\textbf {\bibinfo
  {volume} {2}},\ \bibinfo {pages} {044007} (\bibinfo {year}
  {2017})}\BibitemShut {NoStop}%
\bibitem [{\citenamefont {Majland}, \citenamefont {Christensen},\ and\
  \citenamefont {Zinner}(2020)}]{PhysRevB.101.184510}%
  \BibitemOpen
  \bibfield  {author} {\bibinfo {author} {\bibfnamefont {M.}~\bibnamefont
  {Majland}}, \bibinfo {author} {\bibfnamefont {K.~S.}\ \bibnamefont
  {Christensen}},\ and\ \bibinfo {author} {\bibfnamefont {N.~T.}\ \bibnamefont
  {Zinner}},\ }\bibfield  {title} {\enquote {\bibinfo {title} {Quantum thermal
  transistor in superconducting circuits},}\ }\href
  {https://doi.org/10.1103/PhysRevB.101.184510} {\bibfield  {journal} {\bibinfo
   {journal} {Phys. Rev. B}\ }\textbf {\bibinfo {volume} {101}},\ \bibinfo
  {pages} {184510} (\bibinfo {year} {2020})}\BibitemShut {NoStop}%
\bibitem [{\citenamefont {Xu}, \citenamefont {Stockburger},\ and\ \citenamefont
  {Ankerhold}(2021)}]{PhysRevB.103.104304}%
  \BibitemOpen
  \bibfield  {author} {\bibinfo {author} {\bibfnamefont {M.}~\bibnamefont
  {Xu}}, \bibinfo {author} {\bibfnamefont {J.~T.}\ \bibnamefont
  {Stockburger}},\ and\ \bibinfo {author} {\bibfnamefont {J.}~\bibnamefont
  {Ankerhold}},\ }\bibfield  {title} {\enquote {\bibinfo {title} {Heat
  transport through a superconducting artificial atom},}\ }\href
  {https://doi.org/10.1103/PhysRevB.103.104304} {\bibfield  {journal} {\bibinfo
   {journal} {Phys. Rev. B}\ }\textbf {\bibinfo {volume} {103}},\ \bibinfo
  {pages} {104304} (\bibinfo {year} {2021})}\BibitemShut {NoStop}%
\bibitem [{\citenamefont {Tian}, \citenamefont {Allman},\ and\ \citenamefont
  {Simmonds}(2008)}]{Tian_2008}%
  \BibitemOpen
  \bibfield  {author} {\bibinfo {author} {\bibfnamefont {L.}~\bibnamefont
  {Tian}}, \bibinfo {author} {\bibfnamefont {M.~S.}\ \bibnamefont {Allman}},\
  and\ \bibinfo {author} {\bibfnamefont {R.~W.}\ \bibnamefont {Simmonds}},\
  }\bibfield  {title} {\enquote {\bibinfo {title} {Parametric coupling between
  macroscopic quantum resonators},}\ }\href
  {https://doi.org/10.1088/1367-2630/10/11/115001} {\bibfield  {journal}
  {\bibinfo  {journal} {New J. Phys.}\ }\textbf {\bibinfo {volume} {10}},\
  \bibinfo {pages} {115001} (\bibinfo {year} {2008})}\BibitemShut {NoStop}%
\bibitem [{\citenamefont {Allman}\ \emph {et~al.}(2010)\citenamefont {Allman},
  \citenamefont {Altomare}, \citenamefont {Whittaker}, \citenamefont {Cicak},
  \citenamefont {Li}, \citenamefont {Sirois}, \citenamefont {Strong},
  \citenamefont {Teufel},\ and\ \citenamefont
  {Simmonds}}]{PhysRevLett.104.177004}%
  \BibitemOpen
  \bibfield  {author} {\bibinfo {author} {\bibfnamefont {M.~S.}\ \bibnamefont
  {Allman}}, \bibinfo {author} {\bibfnamefont {F.}~\bibnamefont {Altomare}},
  \bibinfo {author} {\bibfnamefont {J.~D.}\ \bibnamefont {Whittaker}}, \bibinfo
  {author} {\bibfnamefont {K.}~\bibnamefont {Cicak}}, \bibinfo {author}
  {\bibfnamefont {D.}~\bibnamefont {Li}}, \bibinfo {author} {\bibfnamefont
  {A.}~\bibnamefont {Sirois}}, \bibinfo {author} {\bibfnamefont
  {J.}~\bibnamefont {Strong}}, \bibinfo {author} {\bibfnamefont {J.~D.}\
  \bibnamefont {Teufel}},\ and\ \bibinfo {author} {\bibfnamefont {R.~W.}\
  \bibnamefont {Simmonds}},\ }\bibfield  {title} {\enquote {\bibinfo {title}
  {rf-squid-mediated coherent tunable coupling between a superconducting phase
  qubit and a lumped-element resonator},}\ }\href
  {https://doi.org/10.1103/PhysRevLett.104.177004} {\bibfield  {journal}
  {\bibinfo  {journal} {Phys. Rev. Lett.}\ }\textbf {\bibinfo {volume} {104}},\
  \bibinfo {pages} {177004} (\bibinfo {year} {2010})}\BibitemShut {NoStop}%
\bibitem [{\citenamefont {Bialczak}\ \emph {et~al.}(2011)\citenamefont
  {Bialczak}, \citenamefont {Ansmann}, \citenamefont {Hofheinz}, \citenamefont
  {Lenander}, \citenamefont {Lucero}, \citenamefont {Neeley}, \citenamefont
  {O'Connell}, \citenamefont {Sank}, \citenamefont {Wang}, \citenamefont
  {Weides}, \citenamefont {Wenner}, \citenamefont {Yamamoto}, \citenamefont
  {Cleland},\ and\ \citenamefont {Martinis}}]{PhysRevLett.106.060501}%
  \BibitemOpen
  \bibfield  {author} {\bibinfo {author} {\bibfnamefont {R.~C.}\ \bibnamefont
  {Bialczak}}, \bibinfo {author} {\bibfnamefont {M.}~\bibnamefont {Ansmann}},
  \bibinfo {author} {\bibfnamefont {M.}~\bibnamefont {Hofheinz}}, \bibinfo
  {author} {\bibfnamefont {M.}~\bibnamefont {Lenander}}, \bibinfo {author}
  {\bibfnamefont {E.}~\bibnamefont {Lucero}}, \bibinfo {author} {\bibfnamefont
  {M.}~\bibnamefont {Neeley}}, \bibinfo {author} {\bibfnamefont {A.~D.}\
  \bibnamefont {O'Connell}}, \bibinfo {author} {\bibfnamefont {D.}~\bibnamefont
  {Sank}}, \bibinfo {author} {\bibfnamefont {H.}~\bibnamefont {Wang}}, \bibinfo
  {author} {\bibfnamefont {M.}~\bibnamefont {Weides}}, \bibinfo {author}
  {\bibfnamefont {J.}~\bibnamefont {Wenner}}, \bibinfo {author} {\bibfnamefont
  {T.}~\bibnamefont {Yamamoto}}, \bibinfo {author} {\bibfnamefont {A.~N.}\
  \bibnamefont {Cleland}},\ and\ \bibinfo {author} {\bibfnamefont {J.~M.}\
  \bibnamefont {Martinis}},\ }\bibfield  {title} {\enquote {\bibinfo {title}
  {Fast tunable coupler for superconducting qubits},}\ }\href
  {https://doi.org/10.1103/PhysRevLett.106.060501} {\bibfield  {journal}
  {\bibinfo  {journal} {Phys. Rev. Lett.}\ }\textbf {\bibinfo {volume} {106}},\
  \bibinfo {pages} {060501} (\bibinfo {year} {2011})}\BibitemShut {NoStop}%
\bibitem [{\citenamefont {Xu}\ and\ \citenamefont
  {Ansari}(2021)}]{PhysRevApplied.15.064074}%
  \BibitemOpen
  \bibfield  {author} {\bibinfo {author} {\bibfnamefont {X.}~\bibnamefont
  {Xu}}\ and\ \bibinfo {author} {\bibfnamefont {M.}~\bibnamefont {Ansari}},\
  }\bibfield  {title} {\enquote {\bibinfo {title} {{$ZZ$ Freedom in Two-Qubit
  Gates}},}\ }\href {https://doi.org/10.1103/PhysRevApplied.15.064074}
  {\bibfield  {journal} {\bibinfo  {journal} {Phys. Rev. Appl.}\ }\textbf
  {\bibinfo {volume} {15}},\ \bibinfo {pages} {064074} (\bibinfo {year}
  {2021})}\BibitemShut {NoStop}%
\bibitem [{\citenamefont {van~den Brink}, \citenamefont {Berkley},\ and\
  \citenamefont {Yalowsky}(2005)}]{Brink_2005}%
  \BibitemOpen
  \bibfield  {author} {\bibinfo {author} {\bibfnamefont {A.~M.}\ \bibnamefont
  {van~den Brink}}, \bibinfo {author} {\bibfnamefont {A.~J.}\ \bibnamefont
  {Berkley}},\ and\ \bibinfo {author} {\bibfnamefont {M.}~\bibnamefont
  {Yalowsky}},\ }\bibfield  {title} {\enquote {\bibinfo {title} {Mediated
  tunable coupling of flux qubits},}\ }\href
  {https://doi.org/10.1088/1367-2630/7/1/230} {\bibfield  {journal} {\bibinfo
  {journal} {New J. Phys.}\ }\textbf {\bibinfo {volume} {7}},\ \bibinfo {pages}
  {230} (\bibinfo {year} {2005})}\BibitemShut {NoStop}%
\bibitem [{\citenamefont {Poulsen}\ \emph {et~al.}(2022)\citenamefont
  {Poulsen}, \citenamefont {Santos}, \citenamefont {Kristensen},\ and\
  \citenamefont {Zinner}}]{PhysRevA.105.052605}%
  \BibitemOpen
  \bibfield  {author} {\bibinfo {author} {\bibfnamefont {K.}~\bibnamefont
  {Poulsen}}, \bibinfo {author} {\bibfnamefont {A.~C.}\ \bibnamefont {Santos}},
  \bibinfo {author} {\bibfnamefont {L.~B.}\ \bibnamefont {Kristensen}},\ and\
  \bibinfo {author} {\bibfnamefont {N.~T.}\ \bibnamefont {Zinner}},\ }\bibfield
   {title} {\enquote {\bibinfo {title} {Entanglement-enhanced quantum
  rectification},}\ }\href {https://doi.org/10.1103/PhysRevA.105.052605}
  {\bibfield  {journal} {\bibinfo  {journal} {Phys. Rev. A}\ }\textbf {\bibinfo
  {volume} {105}},\ \bibinfo {pages} {052605} (\bibinfo {year}
  {2022})}\BibitemShut {NoStop}%
\bibitem [{\citenamefont {Miyanaga}\ \emph {et~al.}(2021)\citenamefont
  {Miyanaga}, \citenamefont {Tomonaga}, \citenamefont {Ito}, \citenamefont
  {Mukai},\ and\ \citenamefont {Tsai}}]{PhysRevApplied.16.064041}%
  \BibitemOpen
  \bibfield  {author} {\bibinfo {author} {\bibfnamefont {T.}~\bibnamefont
  {Miyanaga}}, \bibinfo {author} {\bibfnamefont {A.}~\bibnamefont {Tomonaga}},
  \bibinfo {author} {\bibfnamefont {H.}~\bibnamefont {Ito}}, \bibinfo {author}
  {\bibfnamefont {H.}~\bibnamefont {Mukai}},\ and\ \bibinfo {author}
  {\bibfnamefont {J.}~\bibnamefont {Tsai}},\ }\bibfield  {title} {\enquote
  {\bibinfo {title} {{Ultrastrong Tunable Coupler Between Superconducting $LC$
  Resonators}},}\ }\href {https://doi.org/10.1103/PhysRevApplied.16.064041}
  {\bibfield  {journal} {\bibinfo  {journal} {Phys. Rev. Appl.}\ }\textbf
  {\bibinfo {volume} {16}},\ \bibinfo {pages} {064041} (\bibinfo {year}
  {2021})}\BibitemShut {NoStop}%
\bibitem [{\citenamefont {Goold}\ \emph {et~al.}(2016)\citenamefont {Goold},
  \citenamefont {Huber}, \citenamefont {Riera}, \citenamefont {del Rio},\ and\
  \citenamefont {Skrzypczyk}}]{Goold_2016}%
  \BibitemOpen
  \bibfield  {author} {\bibinfo {author} {\bibfnamefont {J.}~\bibnamefont
  {Goold}}, \bibinfo {author} {\bibfnamefont {M.}~\bibnamefont {Huber}},
  \bibinfo {author} {\bibfnamefont {A.}~\bibnamefont {Riera}}, \bibinfo
  {author} {\bibfnamefont {L.}~\bibnamefont {del Rio}},\ and\ \bibinfo {author}
  {\bibfnamefont {P.}~\bibnamefont {Skrzypczyk}},\ }\bibfield  {title}
  {\enquote {\bibinfo {title} {The role of quantum information in
  thermodynamics—a topical review},}\ }\href
  {https://doi.org/10.1088/1751-8113/49/14/143001} {\bibfield  {journal}
  {\bibinfo  {journal} {J. Phys. A: Math. Theor.}\ }\textbf {\bibinfo {volume}
  {49}},\ \bibinfo {pages} {143001} (\bibinfo {year} {2016})}\BibitemShut
  {NoStop}%
\bibitem [{\citenamefont {Uzdin}, \citenamefont {Levy},\ and\ \citenamefont
  {Kosloff}(2015)}]{PhysRevX.5.031044}%
  \BibitemOpen
  \bibfield  {author} {\bibinfo {author} {\bibfnamefont {R.}~\bibnamefont
  {Uzdin}}, \bibinfo {author} {\bibfnamefont {A.}~\bibnamefont {Levy}},\ and\
  \bibinfo {author} {\bibfnamefont {R.}~\bibnamefont {Kosloff}},\ }\bibfield
  {title} {\enquote {\bibinfo {title} {Equivalence of quantum heat machines,
  and quantum-thermodynamic signatures},}\ }\href
  {https://doi.org/10.1103/PhysRevX.5.031044} {\bibfield  {journal} {\bibinfo
  {journal} {Phys. Rev. X}\ }\textbf {\bibinfo {volume} {5}},\ \bibinfo {pages}
  {031044} (\bibinfo {year} {2015})}\BibitemShut {NoStop}%
\bibitem [{\citenamefont {Brask}\ \emph {et~al.}(2015)\citenamefont {Brask},
  \citenamefont {Haack}, \citenamefont {Brunner},\ and\ \citenamefont
  {Huber}}]{Bohr_Brask_2015}%
  \BibitemOpen
  \bibfield  {author} {\bibinfo {author} {\bibfnamefont {J.~B.}\ \bibnamefont
  {Brask}}, \bibinfo {author} {\bibfnamefont {G.}~\bibnamefont {Haack}},
  \bibinfo {author} {\bibfnamefont {N.}~\bibnamefont {Brunner}},\ and\ \bibinfo
  {author} {\bibfnamefont {M.}~\bibnamefont {Huber}},\ }\bibfield  {title}
  {\enquote {\bibinfo {title} {Autonomous quantum thermal machine for
  generating steady-state entanglement},}\ }\href
  {https://doi.org/10.1088/1367-2630/17/11/113029} {\bibfield  {journal}
  {\bibinfo  {journal} {New J. Phys.}\ }\textbf {\bibinfo {volume} {17}},\
  \bibinfo {pages} {113029} (\bibinfo {year} {2015})}\BibitemShut {NoStop}%
\bibitem [{\citenamefont {Khandelwal}\ \emph {et~al.}(2020)\citenamefont
  {Khandelwal}, \citenamefont {Palazzo}, \citenamefont {Brunner},\ and\
  \citenamefont {Haack}}]{Khandelwal_2020}%
  \BibitemOpen
  \bibfield  {author} {\bibinfo {author} {\bibfnamefont {S.}~\bibnamefont
  {Khandelwal}}, \bibinfo {author} {\bibfnamefont {N.}~\bibnamefont {Palazzo}},
  \bibinfo {author} {\bibfnamefont {N.}~\bibnamefont {Brunner}},\ and\ \bibinfo
  {author} {\bibfnamefont {G.}~\bibnamefont {Haack}},\ }\bibfield  {title}
  {\enquote {\bibinfo {title} {Critical heat current for operating an
  entanglement engine},}\ }\href {https://doi.org/10.1088/1367-2630/ab9983}
  {\bibfield  {journal} {\bibinfo  {journal} {New J. Phys.}\ }\textbf {\bibinfo
  {volume} {22}},\ \bibinfo {pages} {073039} (\bibinfo {year}
  {2020})}\BibitemShut {NoStop}%
\bibitem [{\citenamefont {Hammam}\ \emph {et~al.}(2022)\citenamefont {Hammam},
  \citenamefont {Leitch}, \citenamefont {Hassouni},\ and\ \citenamefont
  {Chiara}}]{Hammam_2022}%
  \BibitemOpen
  \bibfield  {author} {\bibinfo {author} {\bibfnamefont {K.}~\bibnamefont
  {Hammam}}, \bibinfo {author} {\bibfnamefont {H.}~\bibnamefont {Leitch}},
  \bibinfo {author} {\bibfnamefont {Y.}~\bibnamefont {Hassouni}},\ and\
  \bibinfo {author} {\bibfnamefont {G.~D.}\ \bibnamefont {Chiara}},\ }\bibfield
   {title} {\enquote {\bibinfo {title} {Exploiting coherence for quantum
  thermodynamic advantage},}\ }\href {https://doi.org/10.1088/1367-2630/aca49b}
  {\bibfield  {journal} {\bibinfo  {journal} {New J. Phys.}\ }\textbf {\bibinfo
  {volume} {24}},\ \bibinfo {pages} {113053} (\bibinfo {year}
  {2022})}\BibitemShut {NoStop}%
\bibitem [{\citenamefont {Richer}\ \emph {et~al.}(2017)\citenamefont {Richer},
  \citenamefont {Maleeva}, \citenamefont {Skacel}, \citenamefont {Pop},\ and\
  \citenamefont {DiVincenzo}}]{PhysRevB.96.174520}%
  \BibitemOpen
  \bibfield  {author} {\bibinfo {author} {\bibfnamefont {S.}~\bibnamefont
  {Richer}}, \bibinfo {author} {\bibfnamefont {N.}~\bibnamefont {Maleeva}},
  \bibinfo {author} {\bibfnamefont {S.~T.}\ \bibnamefont {Skacel}}, \bibinfo
  {author} {\bibfnamefont {I.~M.}\ \bibnamefont {Pop}},\ and\ \bibinfo {author}
  {\bibfnamefont {D.}~\bibnamefont {DiVincenzo}},\ }\bibfield  {title}
  {\enquote {\bibinfo {title} {Inductively shunted transmon qubit with tunable
  transverse and longitudinal coupling},}\ }\href
  {https://doi.org/10.1103/PhysRevB.96.174520} {\bibfield  {journal} {\bibinfo
  {journal} {Phys. Rev. B}\ }\textbf {\bibinfo {volume} {96}},\ \bibinfo
  {pages} {174520} (\bibinfo {year} {2017})}\BibitemShut {NoStop}%
\bibitem [{\citenamefont {Sandberg}\ \emph {et~al.}(2008)\citenamefont
  {Sandberg}, \citenamefont {Wilson}, \citenamefont {Persson}, \citenamefont
  {Bauch}, \citenamefont {Johansson}, \citenamefont {Shumeiko}, \citenamefont
  {Duty},\ and\ \citenamefont {Delsing}}]{sandberg2008tuning}%
  \BibitemOpen
  \bibfield  {author} {\bibinfo {author} {\bibfnamefont {M.}~\bibnamefont
  {Sandberg}}, \bibinfo {author} {\bibfnamefont {C.}~\bibnamefont {Wilson}},
  \bibinfo {author} {\bibfnamefont {F.}~\bibnamefont {Persson}}, \bibinfo
  {author} {\bibfnamefont {T.}~\bibnamefont {Bauch}}, \bibinfo {author}
  {\bibfnamefont {G.}~\bibnamefont {Johansson}}, \bibinfo {author}
  {\bibfnamefont {V.}~\bibnamefont {Shumeiko}}, \bibinfo {author}
  {\bibfnamefont {T.}~\bibnamefont {Duty}},\ and\ \bibinfo {author}
  {\bibfnamefont {P.}~\bibnamefont {Delsing}},\ }\bibfield  {title} {\enquote
  {\bibinfo {title} {Tuning the field in a microwave resonator faster than the
  photon lifetime},}\ }\href@noop {} {\bibfield  {journal} {\bibinfo  {journal}
  {Appl. Phys. Lett.}\ }\textbf {\bibinfo {volume} {92}},\ \bibinfo {pages}
  {203501} (\bibinfo {year} {2008})}\BibitemShut {NoStop}%
\bibitem [{\citenamefont {Uhl}\ \emph {et~al.}(2023)\citenamefont {Uhl},
  \citenamefont {Hackenbeck}, \citenamefont {Füger}, \citenamefont {Kleiner},
  \citenamefont {Koelle},\ and\ \citenamefont {Bothner}}]{Uhl}%
  \BibitemOpen
  \bibfield  {author} {\bibinfo {author} {\bibfnamefont {K.}~\bibnamefont
  {Uhl}}, \bibinfo {author} {\bibfnamefont {D.}~\bibnamefont {Hackenbeck}},
  \bibinfo {author} {\bibfnamefont {C.}~\bibnamefont {Füger}}, \bibinfo
  {author} {\bibfnamefont {R.}~\bibnamefont {Kleiner}}, \bibinfo {author}
  {\bibfnamefont {D.}~\bibnamefont {Koelle}},\ and\ \bibinfo {author}
  {\bibfnamefont {D.}~\bibnamefont {Bothner}},\ }\bibfield  {title} {\enquote
  {\bibinfo {title} {{A flux-tunable YBa2Cu3O7 quantum interference microwave
  circuit}},}\ }\href {https://doi.org/10.1063/5.0146524} {\bibfield  {journal}
  {\bibinfo  {journal} {Appl. Phys. Lett.}\ }\textbf {\bibinfo {volume}
  {122}},\ \bibinfo {pages} {182603} (\bibinfo {year} {2023})}\BibitemShut
  {NoStop}%
\bibitem [{\citenamefont {Ojanen}\ and\ \citenamefont
  {Jauho}(2008)}]{PhysRevLett.100.155902}%
  \BibitemOpen
  \bibfield  {author} {\bibinfo {author} {\bibfnamefont {T.}~\bibnamefont
  {Ojanen}}\ and\ \bibinfo {author} {\bibfnamefont {A.-P.}\ \bibnamefont
  {Jauho}},\ }\bibfield  {title} {\enquote {\bibinfo {title} {Mesoscopic photon
  heat transistor},}\ }\href {https://doi.org/10.1103/PhysRevLett.100.155902}
  {\bibfield  {journal} {\bibinfo  {journal} {Phys. Rev. Lett.}\ }\textbf
  {\bibinfo {volume} {100}},\ \bibinfo {pages} {155902} (\bibinfo {year}
  {2008})}\BibitemShut {NoStop}%
\bibitem [{\citenamefont {Caldeira}\ and\ \citenamefont
  {Leggett}(1983)}]{caldeira1983quantum}%
  \BibitemOpen
  \bibfield  {author} {\bibinfo {author} {\bibfnamefont {A.~O.}\ \bibnamefont
  {Caldeira}}\ and\ \bibinfo {author} {\bibfnamefont {A.~J.}\ \bibnamefont
  {Leggett}},\ }\bibfield  {title} {\enquote {\bibinfo {title} {Quantum
  tunnelling in a dissipative system},}\ }\href@noop {} {\bibfield  {journal}
  {\bibinfo  {journal} {Ann. Phys.}\ }\textbf {\bibinfo {volume} {149}},\
  \bibinfo {pages} {374--456} (\bibinfo {year} {1983})}\BibitemShut {NoStop}%
\bibitem [{\citenamefont {Reuther}\ \emph {et~al.}(2010)\citenamefont
  {Reuther}, \citenamefont {Zueco}, \citenamefont {Deppe}, \citenamefont
  {Hoffmann}, \citenamefont {Menzel}, \citenamefont {Wei\ss{}l}, \citenamefont
  {Mariantoni}, \citenamefont {Kohler}, \citenamefont {Marx}, \citenamefont
  {Solano}, \citenamefont {Gross},\ and\ \citenamefont
  {H\"anggi}}]{PhysRevB.81.144510}%
  \BibitemOpen
  \bibfield  {author} {\bibinfo {author} {\bibfnamefont {G.~M.}\ \bibnamefont
  {Reuther}}, \bibinfo {author} {\bibfnamefont {D.}~\bibnamefont {Zueco}},
  \bibinfo {author} {\bibfnamefont {F.}~\bibnamefont {Deppe}}, \bibinfo
  {author} {\bibfnamefont {E.}~\bibnamefont {Hoffmann}}, \bibinfo {author}
  {\bibfnamefont {E.~P.}\ \bibnamefont {Menzel}}, \bibinfo {author}
  {\bibfnamefont {T.}~\bibnamefont {Wei\ss{}l}}, \bibinfo {author}
  {\bibfnamefont {M.}~\bibnamefont {Mariantoni}}, \bibinfo {author}
  {\bibfnamefont {S.}~\bibnamefont {Kohler}}, \bibinfo {author} {\bibfnamefont
  {A.}~\bibnamefont {Marx}}, \bibinfo {author} {\bibfnamefont {E.}~\bibnamefont
  {Solano}}, \bibinfo {author} {\bibfnamefont {R.}~\bibnamefont {Gross}},\ and\
  \bibinfo {author} {\bibfnamefont {P.}~\bibnamefont {H\"anggi}},\ }\bibfield
  {title} {\enquote {\bibinfo {title} {Two-resonator circuit quantum
  electrodynamics: Dissipative theory},}\ }\href
  {https://doi.org/10.1103/PhysRevB.81.144510} {\bibfield  {journal} {\bibinfo
  {journal} {Phys. Rev. B}\ }\textbf {\bibinfo {volume} {81}},\ \bibinfo
  {pages} {144510} (\bibinfo {year} {2010})}\BibitemShut {NoStop}%
\bibitem [{\citenamefont {Bamba}\ and\ \citenamefont
  {Ogawa}(2014)}]{PhysRevA.89.023817}%
  \BibitemOpen
  \bibfield  {author} {\bibinfo {author} {\bibfnamefont {M.}~\bibnamefont
  {Bamba}}\ and\ \bibinfo {author} {\bibfnamefont {T.}~\bibnamefont {Ogawa}},\
  }\bibfield  {title} {\enquote {\bibinfo {title} {Recipe for the hamiltonian
  of system-environment coupling applicable to the
  ultrastrong-light-matter-interaction regime},}\ }\href
  {https://doi.org/10.1103/PhysRevA.89.023817} {\bibfield  {journal} {\bibinfo
  {journal} {Phys. Rev. A}\ }\textbf {\bibinfo {volume} {89}},\ \bibinfo
  {pages} {023817} (\bibinfo {year} {2014})}\BibitemShut {NoStop}%
\bibitem [{\citenamefont {Niskanen}, \citenamefont {Nakamura},\ and\
  \citenamefont {Pekola}(2007)}]{PhysRevB.76.174523}%
  \BibitemOpen
  \bibfield  {author} {\bibinfo {author} {\bibfnamefont {A.~O.}\ \bibnamefont
  {Niskanen}}, \bibinfo {author} {\bibfnamefont {Y.}~\bibnamefont {Nakamura}},\
  and\ \bibinfo {author} {\bibfnamefont {J.~P.}\ \bibnamefont {Pekola}},\
  }\bibfield  {title} {\enquote {\bibinfo {title} {Information entropic
  superconducting microcooler},}\ }\href
  {https://doi.org/10.1103/PhysRevB.76.174523} {\bibfield  {journal} {\bibinfo
  {journal} {Phys. Rev. B}\ }\textbf {\bibinfo {volume} {76}},\ \bibinfo
  {pages} {174523} (\bibinfo {year} {2007})}\BibitemShut {NoStop}%
\bibitem [{\citenamefont {Breuer}\ and\ \citenamefont
  {Petruccione}(2002)}]{breuer2002theory}%
  \BibitemOpen
  \bibfield  {author} {\bibinfo {author} {\bibfnamefont {H.-P.}\ \bibnamefont
  {Breuer}}\ and\ \bibinfo {author} {\bibfnamefont {F.}~\bibnamefont
  {Petruccione}},\ }\href@noop {} {\emph {\bibinfo {title} {The theory of open
  quantum systems}}}\ (\bibinfo  {publisher} {Oxford University Press,
  Oxford},\ \bibinfo {year} {2002})\BibitemShut {NoStop}%
\bibitem [{\citenamefont {Vool}\ and\ \citenamefont
  {Devoret}(2017)}]{vool2017introduction}%
  \BibitemOpen
  \bibfield  {author} {\bibinfo {author} {\bibfnamefont {U.}~\bibnamefont
  {Vool}}\ and\ \bibinfo {author} {\bibfnamefont {M.}~\bibnamefont {Devoret}},\
  }\bibfield  {title} {\enquote {\bibinfo {title} {Introduction to quantum
  electromagnetic circuits},}\ }\href@noop {} {\bibfield  {journal} {\bibinfo
  {journal} {International Journal of Circuit Theory and Applications}\
  }\textbf {\bibinfo {volume} {45}},\ \bibinfo {pages} {897--934} (\bibinfo
  {year} {2017})}\BibitemShut {NoStop}%
\bibitem [{\citenamefont {Cattaneo}\ and\ \citenamefont
  {Paraoanu}(2021)}]{https://doi.org/10.1002/qute.202100054}%
  \BibitemOpen
  \bibfield  {author} {\bibinfo {author} {\bibfnamefont {M.}~\bibnamefont
  {Cattaneo}}\ and\ \bibinfo {author} {\bibfnamefont {G.~S.}\ \bibnamefont
  {Paraoanu}},\ }\bibfield  {title} {\enquote {\bibinfo {title} {Engineering
  dissipation with resistive elements in circuit quantum electrodynamics},}\
  }\href {https://doi.org/https://doi.org/10.1002/qute.202100054} {\bibfield
  {journal} {\bibinfo  {journal} {Adv. Quantum Technol.}\ }\textbf {\bibinfo
  {volume} {4}},\ \bibinfo {pages} {2100054} (\bibinfo {year}
  {2021})}\BibitemShut {NoStop}%
\bibitem [{\citenamefont {Vidal}\ and\ \citenamefont
  {Werner}(2002)}]{PhysRevA.65.032314}%
  \BibitemOpen
  \bibfield  {author} {\bibinfo {author} {\bibfnamefont {G.}~\bibnamefont
  {Vidal}}\ and\ \bibinfo {author} {\bibfnamefont {R.~F.}\ \bibnamefont
  {Werner}},\ }\bibfield  {title} {\enquote {\bibinfo {title} {Computable
  measure of entanglement},}\ }\href
  {https://doi.org/10.1103/PhysRevA.65.032314} {\bibfield  {journal} {\bibinfo
  {journal} {Phys. Rev. A}\ }\textbf {\bibinfo {volume} {65}},\ \bibinfo
  {pages} {032314} (\bibinfo {year} {2002})}\BibitemShut {NoStop}%
\bibitem [{\citenamefont {Adesso}, \citenamefont {Serafini},\ and\
  \citenamefont {Illuminati}(2004)}]{PhysRevA.70.022318}%
  \BibitemOpen
  \bibfield  {author} {\bibinfo {author} {\bibfnamefont {G.}~\bibnamefont
  {Adesso}}, \bibinfo {author} {\bibfnamefont {A.}~\bibnamefont {Serafini}},\
  and\ \bibinfo {author} {\bibfnamefont {F.}~\bibnamefont {Illuminati}},\
  }\bibfield  {title} {\enquote {\bibinfo {title} {Extremal entanglement and
  mixedness in continuous variable systems},}\ }\href
  {https://doi.org/10.1103/PhysRevA.70.022318} {\bibfield  {journal} {\bibinfo
  {journal} {Phys. Rev. A}\ }\textbf {\bibinfo {volume} {70}},\ \bibinfo
  {pages} {022318} (\bibinfo {year} {2004})}\BibitemShut {NoStop}%
\bibitem [{\citenamefont {Naseem}, \citenamefont {Xuereb},\ and\ \citenamefont
  {M\"ustecapl\ifmmode \imath \else \i \fi{}o\ifmmode~\breve{g}\else
  \u{g}\fi{}lu}(2018)}]{PhysRevA.98.052123}%
  \BibitemOpen
  \bibfield  {author} {\bibinfo {author} {\bibfnamefont {M.~T.}\ \bibnamefont
  {Naseem}}, \bibinfo {author} {\bibfnamefont {A.}~\bibnamefont {Xuereb}},\
  and\ \bibinfo {author} {\bibfnamefont {O.~E.}\ \bibnamefont
  {M\"ustecapl\ifmmode \imath \else \i \fi{}o\ifmmode~\breve{g}\else
  \u{g}\fi{}lu}},\ }\bibfield  {title} {\enquote {\bibinfo {title}
  {Thermodynamic consistency of the optomechanical master equation},}\ }\href
  {https://doi.org/10.1103/PhysRevA.98.052123} {\bibfield  {journal} {\bibinfo
  {journal} {Phys. Rev. A}\ }\textbf {\bibinfo {volume} {98}},\ \bibinfo
  {pages} {052123} (\bibinfo {year} {2018})}\BibitemShut {NoStop}%
\bibitem [{\citenamefont {Giazotto}\ \emph {et~al.}(2006)\citenamefont
  {Giazotto}, \citenamefont {Heikkil\"a}, \citenamefont {Luukanen},
  \citenamefont {Savin},\ and\ \citenamefont {Pekola}}]{RevModPhys.78.217}%
  \BibitemOpen
  \bibfield  {author} {\bibinfo {author} {\bibfnamefont {F.}~\bibnamefont
  {Giazotto}}, \bibinfo {author} {\bibfnamefont {T.~T.}\ \bibnamefont
  {Heikkil\"a}}, \bibinfo {author} {\bibfnamefont {A.}~\bibnamefont
  {Luukanen}}, \bibinfo {author} {\bibfnamefont {A.~M.}\ \bibnamefont
  {Savin}},\ and\ \bibinfo {author} {\bibfnamefont {J.~P.}\ \bibnamefont
  {Pekola}},\ }\bibfield  {title} {\enquote {\bibinfo {title} {Opportunities
  for mesoscopics in thermometry and refrigeration: Physics and
  applications},}\ }\href {https://doi.org/10.1103/RevModPhys.78.217}
  {\bibfield  {journal} {\bibinfo  {journal} {Rev. Mod. Phys.}\ }\textbf
  {\bibinfo {volume} {78}},\ \bibinfo {pages} {217--274} (\bibinfo {year}
  {2006})}\BibitemShut {NoStop}%
\bibitem [{\citenamefont {Guthrie}\ \emph {et~al.}(2022)\citenamefont
  {Guthrie}, \citenamefont {Satrya}, \citenamefont {Chang}, \citenamefont
  {Menczel}, \citenamefont {Nori},\ and\ \citenamefont
  {Pekola}}]{PhysRevApplied.17.064022}%
  \BibitemOpen
  \bibfield  {author} {\bibinfo {author} {\bibfnamefont {A.}~\bibnamefont
  {Guthrie}}, \bibinfo {author} {\bibfnamefont {C.~D.}\ \bibnamefont {Satrya}},
  \bibinfo {author} {\bibfnamefont {Y.-C.}\ \bibnamefont {Chang}}, \bibinfo
  {author} {\bibfnamefont {P.}~\bibnamefont {Menczel}}, \bibinfo {author}
  {\bibfnamefont {F.}~\bibnamefont {Nori}},\ and\ \bibinfo {author}
  {\bibfnamefont {J.~P.}\ \bibnamefont {Pekola}},\ }\bibfield  {title}
  {\enquote {\bibinfo {title} {Cooper-pair box coupled to two resonators: An
  architecture for a quantum refrigerator},}\ }\href
  {https://doi.org/10.1103/PhysRevApplied.17.064022} {\bibfield  {journal}
  {\bibinfo  {journal} {Phys. Rev. Appl.}\ }\textbf {\bibinfo {volume} {17}},\
  \bibinfo {pages} {064022} (\bibinfo {year} {2022})}\BibitemShut {NoStop}%
\bibitem [{\citenamefont {Senior}\ \emph {et~al.}(2020)\citenamefont {Senior},
  \citenamefont {Gubaydullin}, \citenamefont {Karimi}, \citenamefont
  {Peltonen}, \citenamefont {Ankerhold},\ and\ \citenamefont
  {Pekola}}]{senior2020heat}%
  \BibitemOpen
  \bibfield  {author} {\bibinfo {author} {\bibfnamefont {J.}~\bibnamefont
  {Senior}}, \bibinfo {author} {\bibfnamefont {A.}~\bibnamefont {Gubaydullin}},
  \bibinfo {author} {\bibfnamefont {B.}~\bibnamefont {Karimi}}, \bibinfo
  {author} {\bibfnamefont {J.~T.}\ \bibnamefont {Peltonen}}, \bibinfo {author}
  {\bibfnamefont {J.}~\bibnamefont {Ankerhold}},\ and\ \bibinfo {author}
  {\bibfnamefont {J.~P.}\ \bibnamefont {Pekola}},\ }\bibfield  {title}
  {\enquote {\bibinfo {title} {Heat rectification via a superconducting
  artificial atom},}\ }\href@noop {} {\bibfield  {journal} {\bibinfo  {journal}
  {Communications Physics}\ }\textbf {\bibinfo {volume} {3}},\ \bibinfo {pages}
  {40} (\bibinfo {year} {2020})}\BibitemShut {NoStop}%
\bibitem [{\citenamefont {Barzanjeh}\ \emph {et~al.}(2019)\citenamefont
  {Barzanjeh}, \citenamefont {Redchenko}, \citenamefont {Peruzzo},
  \citenamefont {Wulf}, \citenamefont {Lewis}, \citenamefont {Arnold},\ and\
  \citenamefont {Fink}}]{barzanjeh2019stationary}%
  \BibitemOpen
  \bibfield  {author} {\bibinfo {author} {\bibfnamefont {S.}~\bibnamefont
  {Barzanjeh}}, \bibinfo {author} {\bibfnamefont {E.}~\bibnamefont
  {Redchenko}}, \bibinfo {author} {\bibfnamefont {M.}~\bibnamefont {Peruzzo}},
  \bibinfo {author} {\bibfnamefont {M.}~\bibnamefont {Wulf}}, \bibinfo {author}
  {\bibfnamefont {D.}~\bibnamefont {Lewis}}, \bibinfo {author} {\bibfnamefont
  {G.}~\bibnamefont {Arnold}},\ and\ \bibinfo {author} {\bibfnamefont {J.~M.}\
  \bibnamefont {Fink}},\ }\bibfield  {title} {\enquote {\bibinfo {title}
  {Stationary entangled radiation from micromechanical motion},}\ }\href@noop
  {} {\bibfield  {journal} {\bibinfo  {journal} {Nature}\ }\textbf {\bibinfo
  {volume} {570}},\ \bibinfo {pages} {480--483} (\bibinfo {year}
  {2019})}\BibitemShut {NoStop}%
\bibitem [{\citenamefont {Chen}\ \emph {et~al.}(2020)\citenamefont {Chen},
  \citenamefont {Rossi}, \citenamefont {Mason},\ and\ \citenamefont
  {Schliesser}}]{chen2020entanglement}%
  \BibitemOpen
  \bibfield  {author} {\bibinfo {author} {\bibfnamefont {J.}~\bibnamefont
  {Chen}}, \bibinfo {author} {\bibfnamefont {M.}~\bibnamefont {Rossi}},
  \bibinfo {author} {\bibfnamefont {D.}~\bibnamefont {Mason}},\ and\ \bibinfo
  {author} {\bibfnamefont {A.}~\bibnamefont {Schliesser}},\ }\bibfield  {title}
  {\enquote {\bibinfo {title} {Entanglement of propagating optical modes via a
  mechanical interface},}\ }\href@noop {} {\bibfield  {journal} {\bibinfo
  {journal} {Nat. Commun.}\ }\textbf {\bibinfo {volume} {11}},\ \bibinfo
  {pages} {943} (\bibinfo {year} {2020})}\BibitemShut {NoStop}%
\bibitem [{\citenamefont {Kotler}\ \emph {et~al.}(2021)\citenamefont {Kotler},
  \citenamefont {Peterson}, \citenamefont {Shojaee}, \citenamefont {Lecocq},
  \citenamefont {Cicak}, \citenamefont {Kwiatkowski}, \citenamefont {Geller},
  \citenamefont {Glancy}, \citenamefont {Knill}, \citenamefont {Simmonds} \emph
  {et~al.}}]{kotler2021direct}%
  \BibitemOpen
  \bibfield  {author} {\bibinfo {author} {\bibfnamefont {S.}~\bibnamefont
  {Kotler}}, \bibinfo {author} {\bibfnamefont {G.~A.}\ \bibnamefont
  {Peterson}}, \bibinfo {author} {\bibfnamefont {E.}~\bibnamefont {Shojaee}},
  \bibinfo {author} {\bibfnamefont {F.}~\bibnamefont {Lecocq}}, \bibinfo
  {author} {\bibfnamefont {K.}~\bibnamefont {Cicak}}, \bibinfo {author}
  {\bibfnamefont {A.}~\bibnamefont {Kwiatkowski}}, \bibinfo {author}
  {\bibfnamefont {S.}~\bibnamefont {Geller}}, \bibinfo {author} {\bibfnamefont
  {S.}~\bibnamefont {Glancy}}, \bibinfo {author} {\bibfnamefont
  {E.}~\bibnamefont {Knill}}, \bibinfo {author} {\bibfnamefont {R.~W.}\
  \bibnamefont {Simmonds}}, \emph {et~al.},\ }\bibfield  {title} {\enquote
  {\bibinfo {title} {Direct observation of deterministic macroscopic
  entanglement},}\ }\href@noop {} {\bibfield  {journal} {\bibinfo  {journal}
  {Science}\ }\textbf {\bibinfo {volume} {372}},\ \bibinfo {pages} {622--625}
  (\bibinfo {year} {2021})}\BibitemShut {NoStop}%
\end{thebibliography}%


\providecommand{\noopsort}[1]{}\providecommand{\singleletter}[1]{#1}%
\begin{thebibliography}{1}%
\makeatletter
\providecommand \@ifxundefined [1]{%
 \@ifx{#1\undefined}
}%
\providecommand \@ifnum [1]{%
 \ifnum #1\expandafter \@firstoftwo
 \else \expandafter \@secondoftwo
 \fi
}%
\providecommand \@ifx [1]{%
 \ifx #1\expandafter \@firstoftwo
 \else \expandafter \@secondoftwo
 \fi
}%
\providecommand \natexlab [1]{#1}%
\providecommand \enquote  [1]{``#1''}%
\providecommand \bibnamefont  [1]{#1}%
\providecommand \bibfnamefont [1]{#1}%
\providecommand \citenamefont [1]{#1}%
\providecommand \href@noop [0]{\@secondoftwo}%
\providecommand \href [0]{\begingroup \@sanitize@url \@href}%
\providecommand \@href[1]{\@@startlink{#1}\@@href}%
\providecommand \@@href[1]{\endgroup#1\@@endlink}%
\providecommand \@sanitize@url [0]{\catcode `\\12\catcode `\$12\catcode
  `\&12\catcode `\#12\catcode `\^12\catcode `\_12\catcode `\%12\relax}%
\providecommand \@@startlink[1]{}%
\providecommand \@@endlink[0]{}%
\providecommand \url  [0]{\begingroup\@sanitize@url \@url }%
\providecommand \@url [1]{\endgroup\@href {#1}{\urlprefix }}%
\providecommand \urlprefix  [0]{URL }%
\providecommand \Eprint [0]{\href }%
\providecommand \doibase [0]{https://doi.org/}%
\providecommand \selectlanguage [0]{\@gobble}%
\providecommand \bibinfo  [0]{\@secondoftwo}%
\providecommand \bibfield  [0]{\@secondoftwo}%
\providecommand \translation [1]{[#1]}%
\providecommand \BibitemOpen [0]{}%
\providecommand \bibitemStop [0]{}%
\providecommand \bibitemNoStop [0]{.\EOS\space}%
\providecommand \EOS [0]{\spacefactor3000\relax}%
\providecommand \BibitemShut  [1]{\csname bibitem#1\endcsname}%
\let\auto@bib@innerbib\@empty
\bibitem [{\citenamefont {Hopfield}(1958)}]{PhysRev.112.1555}%
  \BibitemOpen
  \bibfield  {author} {\bibinfo {author} {\bibfnamefont {J.~J.}\ \bibnamefont
  {Hopfield}},\ }\bibfield  {title} {\enquote {\bibinfo {title} {Theory of the
  contribution of excitons to the complex dielectric constant of crystals},}\
  }\href {https://doi.org/10.1103/PhysRev.112.1555} {\bibfield  {journal}
  {\bibinfo  {journal} {Phys. Rev.}\ }\textbf {\bibinfo {volume} {112}},\
  \bibinfo {pages} {1555--1567} (\bibinfo {year} {1958})}\BibitemShut {NoStop}%
\end{thebibliography}%

\end{document}


\title{Supplemental Material: Quantum heat valve and entanglement in superconducting $L C$ resonators}

\author{Yu-qiang Liu}
\affiliation{School of Physics, Dalian University of Technology, Dalian 116024, China}
\author{ Yi-jia Yang}
\affiliation{School of Physics, Dalian University of Technology, Dalian 116024, China}
\author{Ting-ting Ma}
\affiliation{School of Physics, Dalian University of Technology, Dalian 116024, China}
\author{Chang-shui Yu}
\email{ycs@dlut.edu.cn}

\affiliation{School of Physics, Dalian University of Technology, Dalian 116024, China}

\date{\today}


\maketitle

\section{The Hamiltonian diagonalization of coupled resonators} \label{Appendix A}
The Hamiltonian (1) in the main text can be exactly diagonalized by Hopfied transformation \cite{PhysRev.112.1555}  as 
\begin{equation} \tag{S1}
H_{\mathrm{Hop}} =\sum_{j=\pm} \omega_{j} p_j^{\dagger} p_j,
\end{equation}
where $p_{\pm}=w_{\pm} a+x_{\pm} b+y_{\pm} a^{\dagger}+z_{\pm} b^{\dagger}$ satisfy the relation $[p_{i}, p^{\dagger}_{j}]=\delta_{i, j}$ with the normalized coefficients $w_{\pm}$, $y_{\pm}$, $x_{\pm}$ and $z_{\pm}$, and two eigenmodes are  
\begin{equation} \tag{S2}
\omega_{\pm}=\sqrt{\frac{\omega_{L}^2+\omega^{2}_{R} \pm \sqrt{\left(\omega_{L}^2-\omega^{2}_{R}\right)^2+16  g^2\omega_{L} \omega_{R}}}{2}},
\end{equation}
 determined by the following equation 
\begin{align} \tag{S3}
\left(\begin{array}{cccc}
\omega_{L} & g & 0 & g \\
g & \omega_{R} & g & 0 \\
0 & -g & -\omega_{L} & g \\
-g & 0 & -g & -\omega_{R}
\end{array}\right)\left[\begin{array}{c}
w_{j} \\
x_{j} \\
y_{j} \\
z_{j}
\end{array}\right]=\omega_{\pm} \left[\begin{array}{c}
w_{j} \\
x_{j} \\
y_{j} \\
z_{j}
\end{array}\right].
\end{align}
Similarly, the operators $a$, and $b$ can also be denoted by the polariton operators as
\begin{align}  \tag{S4} \label{original operator}
\begin{aligned}
&a=w_{+}^* p_{+}+w_{-}^* p_{-}-y_{+} p_{+}^{\dagger}-y_{-} p_{-}^{\dagger}, \\
&b=x_{+}^* p_{+}+x_{-}^* p_{-}-z_{+} p_{+}^{\dagger}-z_{-} \hat{p}_{-}^{\dagger}.
\end{aligned}
\end{align}
The corresponding coefficients can be solved as 
\begin{align} \tag{S5}
\left.\left(\begin{array}{l}
w_{\pm} \\
x_{\pm} \\
y_{\pm} \\
z_{\pm}
\end{array}\right)=\right. {\frac{1}{N}}
\left(\begin{array}{c}
{-\frac{(\omega_{\pm}+\omega_{L})(\omega_{\pm}+\omega_{R})}{2 g \omega_{L}}} \\
-1+\frac{(-\omega^2_{\pm}+\omega^2_{L})(\omega_{\pm}+\omega_{R})}{2 g^2 \omega_{L}} \\
\frac{(\omega_{\pm}-\omega_{L})(\omega_{\pm}+\omega_{R})}{2 g \omega_{L}} \\
1
\end{array}\right),
\end{align}
where $N$ is the normalized coefficient. 
Substituting Eq. (\ref{original operator}) into Eq. (2) in the main text, the system-reservoir interaction can be rewritten as
\begin{equation} \tag{S6}
H_{I} =(B_{-}^{\dagger} p_{-}+p_{-}^{\dagger} B_{-})+ (B_{+}^{\dagger} p_{+}+p_{+}^{\dagger} B_{+}),
\end{equation}
where
$B_j=W_{j}(B_L+B_L^{\dagger})+X_{j}( B_R + B_R^{\dagger})
$ with $W_{j}=(w_j-y_j)$ and $X_{j}= (x_j -z_j)$.

\section{The derivation of the master equations and system dynamics}
\label{Appendix B}
The master equation of the evolution system under the Born-Markov approximation can be represented as 
\begin{equation} \tag{S7}
\begin{split}
\dot{\rho}(t)=&\int_0^{\infty} d s  \operatorname{Tr}\left[ - H_I(t) H_I(t-s) \rho(t) \otimes \rho_E\right.
+ H_I(t) \rho(t) \otimes \rho_E H_I(t-s) 
\\&+ H_I(t-s) \rho(t) \otimes \rho_E H_I(t) \left.- \rho(t) \otimes \rho_E H_I(t-s) H_I(t)\right].
\end{split}
\end{equation}

\subsection{The derivation of the master equation for coupled bosonic system }
First, we consider the system coupling to two independent reservoirs and the interaction Hamiltonian in the interaction picture read  
$H_{I}(t)=(B^{\dagger}_{-}(t) p_{-} e^{-i \omega_{-} t}+B_{-}(t) p^{\dagger}_{-} e^{i \omega_{-} t}) + (B^{\dagger}_{+}(t) p_{+} e^{-i \omega_{-} t}+B_{+}(t) p^{\dagger}_{+} e^{i \omega_{+} t})
$.
\begin{align} \label{master equation}
\nonumber &\dot{\rho}(t)=\int_0^{\infty} d s \operatorname{Tr} \lbrace -[(B^{\dagger}_{-}(t) p_{-} e^{-i \omega_{-} t}+B_{-}(t) p^{\dagger}_{-} e^{i \omega_{-} t}) + (B^{\dagger}_{+}(t) p_{+} e^{-i \omega_{-} t}+B_{+}(t) p^{\dagger}_{+} e^{i \omega_{+} t})]  \\ &
\nonumber [ (B^{\dagger}_{-}(t-s) p_{-} e^{-i \omega_{-} (t-s)}+B_{-}(t-s) p^{\dagger}_{-} e^{i \omega_{-} (t-s)}) + (B^{\dagger}_{+}(t-s) p_{+} e^{-i \omega_{-} (t-s)}+B_{+}(t-s) p^{\dagger}_{+} e^{i \omega_{+} (t-s)})] \rho(t) \otimes \rho_{E}+\\ &
\nonumber
[(B^{\dagger}_{-}(t) p_{-} e^{-i \omega_{-} t}+B_{-}(t) p^{\dagger}_{-} e^{i \omega_{-} t}) + (B^{\dagger}_{+}(t) p_{+} e^{-i \omega_{-} t}+B_{+}(t) p^{\dagger}_{+} e^{i \omega_{+} t})] \rho(t) \otimes \rho_{E} \\&  \nonumber
[(B^{\dagger}_{-}(t-s) p_{-} e^{-i \omega_{-} (t-s)}+B_{-}(t-s) p^{\dagger}_{-} e^{i \omega_{-} (t-s)}) + (B^{\dagger}_{+}(t-s) p_{+} e^{-i \omega_{-} (t-s)}+B_{+}(t-s) p^{\dagger}_{+} e^{i \omega_{+} (t-s)})] +h.c.\rbrace
\\&  \nonumber
=\int_0^{\infty} d s \operatorname{Tr} \lbrace B^{\dagger}_{-}(t-s) \rho_{E} B_{-}(t) p_{-} \rho p^{\dagger}_{-} e^{i \omega_{-} s}+ B_{-}(t-s) \rho_{E} B^{\dagger}_{-}(t) p^{\dagger}_{-} \rho p_{-} e^{-i \omega_{-} s}+ B^{\dagger}_{+}(t-s) \rho_{E} B_{+}(t) p_{+} \rho p^{\dagger}_{+} e^{i \omega_{-} s} \\& \nonumber
+ B_{+}(t-s) \rho_{E} B^{\dagger}_{+}(t) p^{\dagger}_{+} \rho p_{+} e^{-i \omega_{+} s} -[B^{\dagger}_{-}(t) B_{-}(t-s)\rho_{E} p_{-} p^{\dagger}_{-} \rho e^{-i \omega_{-}}  + 
B_{-}(t) B^{\dagger}_{-}(t-s) \rho_{E} p^{\dagger}_{-} p_{-} \rho e^{i \omega_{-}} \\ &  \nonumber
+B^{\dagger}_{+}(t) B_{+}(t-s)\rho_{E} p_{+} p^{\dagger}_{+} \rho e^{-i \omega_{+}}  + B_{+}(t) B^{\dagger}_{+}(t-s) \rho_{E} p^{\dagger}_{+} p_{+} \rho e^{i \omega_{+}}]+h.c. \rbrace  \\ & \nonumber
=p_{-} \rho p^{\dagger}_{-} \int_0^{\infty} d s \langle B_{-}(t) B^{\dagger}_{-}(t-s) \rangle e^{i \omega_{-} s}  + p^{\dagger}_{-} \rho p_{-} \int_0^{\infty} d s \langle B^{\dagger}_{-}(t) B_{-}(t-s) \rangle  e^{-i \omega_{-} s} +\\    & \nonumber 
p_{+} \rho p^{\dagger}_{+} \int_0^{\infty} d s \langle B_{+}(t) B^{\dagger}_{+}(t-s) \rangle  e^{i \omega_{+} s} +
p^{\dagger}_{+} \rho p_{+} \int_0^{\infty} d s \langle B^{\dagger}_{+}(t) B_{+}(t-s) \rangle e^{-i \omega_{+} s}- \\ & \nonumber p_{-} p^{\dagger}_{-} \rho \int_0^{\infty} d s \langle B^{\dagger}_{-}(t) B_{-}(t-s) \rangle e^{-i \omega_{-}}-
 p^{\dagger}_{-} p_{-} \rho \int_0^{\infty} d s \langle B_{-}(t) B^{\dagger}_{-}(t-s) \rangle e^{i \omega_{-}} -  \\ & 
 p^{\dagger}_{-} p_{-} \rho \int_0^{\infty} d s \langle B^{\dagger}_{+}(t) B_{+}(t-s) \rangle e^{-i \omega_{+} s} -  
 p^{\dagger}_{-} p_{-} \rho \int_0^{\infty} d s \langle B_{+}(t) B^{\dagger}_{+}(t-s) \rangle e^{i \omega_{+} s} +h.c.  \tag{S8}
\end{align}
Here the secular approximation has been used in the derivation, and the reservoir correlation functions can be expressed as
\begin{equation}  \label{correlation function} 
\begin{split}
\Re \int_0^{\infty} d s \langle B_{-}(t) B^{\dagger}_{-}(t-s) \rangle e^{i \omega_{-} s} &=  \lbrace |W_{-}|^{2} G_{L}(-\omega_{-}) + |X_{-}|^{2} G_{R}(-\omega_{-}) \rbrace,  \\ 
\Re \int_0^{\infty} d s \langle B^{\dagger}_{-}(t) B_{-}(t-s) \rangle  e^{-i \omega_{+} s} &=\lbrace |W_{-}|^{2} G_{L}(\omega_{-}) + |X_{-}|^{2} G_{R}(\omega_{-}) \rbrace,  \\  
\Re \int_0^{\infty} d s \langle B_{+}(t) B^{\dagger}_{+}(t-s) \rangle e^{i \omega_{+} s} &=\lbrace |W_{+}|^{2} G_{L}(-\omega_{+})  + |X_{+}|^{2} G_{R+}(\omega_{+}) \rbrace, \\
\Re \int_0^{\infty} d s \langle B^{\dagger}_{+}(t) B_{+}(t-s) \rangle e^{-i \omega_{+} s} &=\lbrace |W_{+}|^{2} G_{L}(\omega_{+})  + |X_{+}|^{2} G_{R}(\omega_{+}) \rbrace,
\end{split} \tag{S9}
\end{equation}
where we have made use of the formula $\int_0^{\infty} d s e^{\pm i \epsilon s}=\pi \delta(\epsilon)\pm i \mathrm{P} \frac{1}{\epsilon}$. 

The dynamics of the operator $O$ in the Heisenberg equation for an open system, 
\begin{align}  
\nonumber \frac{d O}{dt}&=\sum_{j=\pm}\lbrace(\Gamma_{L} (-\omega_{j}) |W_{j}|^{2}+\Gamma_{R} (-\omega_{j}) |X_{j}|^{2} )[p^{\dagger}_{j} O p_{j}-\frac{1}{2} \lbrace p^{\dagger}_{j} p_{j}, O \rbrace]+(\Gamma_{R} (\omega_{j}) |X_{j}|^{2} + \Gamma_{L}(\omega_{j}) |W_{j}|^{2}) [p_{j} O p^{\dagger}_{j} -\\& \nonumber \frac{1}{2} \lbrace p_{j} p^{\dagger}_{j}, O \rbrace]\rbrace. \tag{S10}
\end{align}

\section{ Steady-state solution}

As the model is a two-mode Gaussian system, we can introduce an $SU(2)$ algebra, $X=(p^{\dagger}_{+} p_{-}+ p^{\dagger}_{-} p_{+})$, $Y=-i(p^{\dagger}_{+} p_{-}- p^{\dagger}_{-} p_{+})$, $N=(p^{\dagger}_{+} p_{+}+ p^{\dagger}_{-} p_{-})$, $Z=(p^{\dagger}_{+} p_{+}- p^{\dagger}_{-} p_{-})$ to obtain the value of observable of the system. 

\begin{equation} \tag{S11}
\begin{split}
 \frac{d N}{dt}&=-\frac{1}{2}(|W_{+}|^{2} \zeta^{L}_{+}+ |X_{+}|^{2} \zeta^{R}_{+}+ |W_{-}|^{2} \zeta^{L}_{-}+ |X_{-}|^{2} \zeta^{R}_{-}) N -\frac{1}{2}(|W_{+}|^{2} \zeta^{L}_{+}+ |X_{+}|^{2} \zeta^{R}_{+}- |W_{-}|^{2} \zeta^{L}_{-}- |X_{-}|^{2} \zeta^{R}_{-}) Z +\\ &(N^{R}_{+} \zeta^{R}_{+} |X_{+}|^{2}+ N^{L}_{+} \zeta^{L}_{+} |W_{+}|^{2} +N^{R}_{-} \zeta^{R}_{-} |X_{-}|^{2}+ N^{L}_{-} \zeta^{L}_{-} |W_{-}|^{2} ) ,\\
 \frac{d Z}{dt}&=-\frac{1}{2}(|W_{+}|^{2} \zeta^{L}_{+}+ |X_{+}|^{2} \zeta^{R}_{+}+ |W_{-}|^{2} \zeta^{L}_{-}+ |X_{-}|^{2} \zeta^{R}_{-}) Z -\frac{1}{2}(|W_{+}|^{2} \zeta^{L}_{+}+ |X_{+}|^{2} \zeta^{R}_{+}-  \zeta^{L}_{-}- |X_{-}|^{2} \zeta^{R}_{-}) N +\\ &(N^{R}_{+} \zeta^{R}_{+} |X_{+}|^{2}+ N^{L}_{+} \zeta^{L}_{+} |W_{+}|^{2} -N^{R}_{-} \zeta^{R}_{-} |X_{-}|^{2}- N^{L}_{-} \zeta^{L}_{-} |W_{-}|^{2} ),
\end{split}
\end{equation} 
where $\Gamma_{\nu} (-\omega_{+}) -\Gamma_{\nu} (\omega_{+})=\zeta^{\nu}_{+}$, $\Gamma_{\nu} (-\omega_{-}) -\Gamma_{\nu} (\omega_{-})=\zeta^{\nu}_{-}$, $\frac{\Gamma_{\nu}(\omega_{\pm})}{\zeta^{\nu}_{\pm}}=\frac{1}{e^{ \omega_{\pm}/ T_{\nu}}-1}= N^{\nu}_{\pm}$.
 The steady state can be obtained by solving $\frac{d N}{dt}=\frac{dZ}{dt}=0$. Thus the
expectation values of the operators can be expressed as
\begin{equation} \label{N} \tag{S12}
\begin{split}
\left \langle N \right\rangle&=\frac{N^{L}_{-} |W_{-}|^{2} \zeta^{L}_{-}+ N^{R}_{-} |W_{+}|^{2} \zeta^{R}_{-}}{|W_{-}|^{2} \zeta^{L}_{-}+ |X_{-}|^{2} \zeta^{R}_{-}}+\frac{N^{L}_{+} |W_{+}|^{2} \zeta^{L}_{+}+ N^{R}_{+} |W_{+}|^{2} \zeta^{R}_{-}}{|W_{+}|^{2} \zeta^{L}_{-}+ |X_{+}|^{2} \zeta^{R}_{+}},\\ 
\left \langle Z \right\rangle&=-\frac{N^{L}_{-} |W_{-}|^{2} \zeta^{L}_{-}+ N^{R}_{-} |W_{+}|^{2} \zeta^{R}_{-}}{|W_{-}|^{2} \zeta^{L}_{-}+ |X_{-}|^{2} \zeta^{R}_{-}}+\frac{N^{L}_{+} |W_{+}|^{2} \zeta^{L}_{+}+ N^{R}_{+} |W_{+}|^{2} \zeta^{R}_{-}}{|W_{+}|^{2} \zeta^{L}_{-}+ |X_{+}|^{2} \zeta^{R}_{+}}.
\end{split}
\end{equation}
The heat currents can be obtained from the above two equations as in Eq. (5) in the main text.

The equations of motion for each of
the second moment can be written as
\begin{equation} \tag{S13}
\begin{split}
\frac{\partial}{\partial t}\left\langle d_+^2\right\rangle & =-(\zeta^{L}_{+} \vert W_{+}\vert^2+\zeta^{R}_{+} \vert X_{+}\vert^2) \left\langle d_+^2\right\rangle+\zeta^{L}_{+} \vert W_{+}\vert^2 (N^{L}_{+}+\frac{1}{2}) +\zeta^{R}_{+} \vert X_{+}\vert^2 (N^{R}_{+}+\frac{1}{2}),\\ 
\frac{\partial}{\partial t} \left\langle d_+ f_+\right\rangle & =-(\zeta^{L}_{+} \vert W_{+}\vert^2+\zeta^{R}_{+} \vert X_{+}\vert^2) \left\langle d_+ f_+\right\rangle+\frac{i (\zeta^{L}_{+} \vert W_{+}\vert^2+\zeta^{R}_{+} \vert X_{+}\vert^2)}{2},\\ 
\frac{\partial}{\partial t}\left\langle f_+^2\right\rangle & =-(\zeta^{L}_{+} \vert W_{+}\vert^2+\zeta^{R}_{+} \vert X_{+}\vert^2) \left\langle f_+^2\right\rangle+\zeta^{L}_{+} \vert W_{+}\vert^2 (N^{L}_{+}+\frac{1}{2}) +\zeta^{R}_{+} \vert X_{+}\vert^2 (N^{R}_{+}+\frac{1}{2}), \\
\frac{\partial}{\partial t}\left\langle d_-^2\right\rangle & =-(\zeta^{L}_{-} \vert W_{-}\vert^2+\zeta^{R}_{-} \vert X_{-}\vert^2) \left\langle d_-^2\right\rangle+\zeta^{L}_{-} \vert W_{-}\vert^2 (N^{L}_{-}+\frac{1}{2}) +\zeta^{R}_{-} \vert X_{-}\vert^2 (N^{R}_{-}+\frac{1}{2}), \\
\frac{\partial}{\partial t}\left\langle d_- f_-\right\rangle & =-(\zeta^{L}_{-} \vert W_{-}\vert^2+\zeta^{R}_{-} \vert X_{-}\vert^2) \left\langle d_- f_-\right\rangle+\frac{i (\zeta^{L}_{-} \vert W_{-}\vert^2+\zeta^{R}_{-} \vert X_{-}\vert^2)}{2}, \\
\frac{\partial}{\partial t}\left\langle f_-^2\right\rangle & =-(\zeta^{L}_{-} \vert W_{-}\vert^2+\zeta^{R}_{-} \vert X_{-}\vert^2) \left\langle f_-^2\right\rangle+\zeta^{L}_{-} \vert W_{-}\vert^2 (N^{L}_{-}+\frac{1}{2}) +\zeta^{R}_{-} \vert X_{-}\vert^2 (N^{R}_{-}+\frac{1}{2}), \\
\frac{\partial}{\partial t}\left\langle d_- d_+\right\rangle & =-(\zeta^{L}_{+} \vert W_{+}\vert^2+\zeta^{R}_{+} \vert X_{+}\vert^2+\zeta^{L}_{-} \vert W_{-}\vert^2+\zeta^{R}_{-} \vert X_{-}\vert^2 ) \frac{\langle d_{-} d_{+}\rangle}{2}, \\
\frac{\partial}{\partial t}\left\langle d_+ f_-\right\rangle & =-(\zeta^{L}_{+} \vert W_{+}\vert^2+\zeta^{R}_{+} \vert X_{+}\vert^2+\zeta^{L}_{-} \vert W_{-}\vert^2+\zeta^{R}_{-} \vert X_{-}\vert^2 ) \frac{\langle d_{+} f_{-}\rangle}{2}, \\
\frac{\partial}{\partial t}\left\langle f_+ d_-\right\rangle & = -(\zeta^{L}_{+} \vert W_{+}\vert^2+\zeta^{R}_{+} \vert X_{+}\vert^2+\zeta^{L}_{-} \vert W_{-}\vert^2+\zeta^{R}_{-} \vert X_{-}\vert^2 ) \frac{\langle f_{+} d_{-}\rangle}{2},\\
\frac{\partial}{\partial t}\left\langle f_+ f_-\right\rangle & = -(\zeta^{L}_{+} \vert W_{+}\vert^2+\zeta^{R}_{+} \vert X_{+}\vert^2+\zeta^{L}_{-} \vert W_{-}\vert^2+\zeta^{R}_{-} \vert X_{-}\vert^2 ) \frac{\langle f_{+} f_{-}\rangle}{2}.
\end{split}
\end{equation}
These equations can be solved as 
\begin{equation} \label{second moments} \tag{S14}
\begin{split}
&\left\langle d_+^2\right\rangle=\left\langle f_+^2\right\rangle=\frac{\zeta^{L}_{+} \vert W_{+}\vert^2 (1+2 N^{L}_{+})+\zeta^{R}_{+} \vert X_{+}\vert^2 (1+2 N^{R}_{+})}{2 (\zeta^{L}_{+} \vert W_{+}\vert^2+\zeta^{R}_{+} \vert X_{+}\vert^2)},\\ 
&\left\langle d_+ f_+\right\rangle=\left\langle d_- f_-\right\rangle=\frac{i}{2}, \\
&\left\langle d_-^2\right\rangle=\left\langle f_-^2\right\rangle=\frac{\zeta^{L}_{-} \vert W_{-}\vert^2 (1+2 N^{L}_{-})+\zeta^{R}_{-} \vert X_{-}\vert^2 (1+2 N^{R}_{-})}{2 (\zeta^{L}_{-} \vert W_{-}\vert^2+\zeta^{R}_{-} \vert X_{-}\vert^2)},\\ 
 &\left\langle d_- d_+\right\rangle=\left\langle d_+ f_-\right\rangle=\left\langle f_+ d_-\right\rangle=\left\langle f_+ f_-\right\rangle=0.
 \end{split}
\end{equation}
Hence the covariance matrix can be expressed as the form in Eq. (7) in the main text. To obtain the covariance matrix $\Gamma^{\prime}$ for the original modes $a$ and $b$, we can use a $S$ matrix to transform $\Gamma$. The $S$ matrix is solved in the following form
\begin{align} \label{S matrix} \tag{S15}
S=\left(\begin{array}{cccc}
\frac{W_{+}+W_{+}^{*}}{2} & i\frac{-W_{+}+W_{+}^{*}}{2} & \frac{W_{-}+W_{-}^{*}}{2} & i\frac{-W_{-}+W_{-}^{*}}{2} \\
i\frac{(w_{+}+y_{+})-(w_{+}^{*}+y_{+}^{*})}{2} & \frac{(w_{+}+y_{+})+(w_{+}^{*}+y_{+}^{*})}{2} & i\frac{(w_{-}+y_{-})-(w_{-}^{*}+y_{-}^{*})}{2} & \frac{(w_{-}+y_{-})+(w_{-}^{*}+y_{-}^{*})}{2} \\
\frac{X_{+}+X_{+}^{*}}{2} & i\frac{-X_{+}+X_{+}^{*}}{2} & \frac{X_{-}+X_{-}^{*}}{2} & i\frac{-X_{-}+X_{-}^{*}}{2} \\
i\frac{(x_{+}+z_{+})-(x_{+}^{*}+z_{+}^{*})}{2} & \frac{(x_{+}+z_{+})+(x_{+}^{*}+z_{+}^{*})}{2} & i\frac{(x_{-}+z_{-})-(x_{-}^{*}+z_{-}^{*})}{2} & \frac{(x_{-}+z_{-})+(x_{-}^{*}+z_{-}^{*})}{2}
\end{array}\right).
\end{align}

\bibliography{ref}